\documentclass[%
 reprint,
superscriptaddress,
 amsmath,amssymb,
 aps,
prb,
]{revtex4-1}

\usepackage{graphicx}
\usepackage{dcolumn}
\usepackage{bm}

\usepackage{color}

\usepackage{color} 

\usepackage{ulem}   

\begin{document}

\preprint{}

\title{Microscopic origin of the Drude-Smith model}

\author{T. L. Cocker}
\email{tyler.cocker@physik.uni-regensburg.de}
\affiliation{%
 Department of Physics, University of Alberta, Edmonton, Alberta T6G 2E1, Canada 
}%
\affiliation{%
 Department of Physics, University of Regensburg, 93040 Regensburg, Germany
}%
\author{D. Baillie}%
\author{M. Buruma}%
\affiliation{%
 Department of Physics, University of Alberta, Edmonton, Alberta T6G 2E1, Canada 
}%
\author{L. V. Titova}%
\affiliation{%
 Department of Physics, Worcester Polytechnic Institute, Worcester, MA 01609-2280, USA
}%
\author{R. D. Sydora}%
\author{F. Marsiglio}%
\author{F. A. Hegmann}%
\email{hegmann@ualberta.ca}
\affiliation{%
 Department of Physics, University of Alberta, Edmonton, Alberta T6G 2E1, Canada 
}%

\date{October 17, 2017} 

\begin{abstract}

The Drude-Smith model has been used extensively in fitting the THz conductivities of nanomaterials with carrier confinement on the mesoscopic scale. Here, we show that the conventional `backscattering' explanation for the suppression of low-frequency conductivities in the Drude-Smith model is not consistent with a confined Drude gas of classical non-interacting electrons and we derive a modified Drude-Smith conductivity formula based on a diffusive restoring current. We perform Monte Carlo simulations of a model system and show that the modified Drude-Smith model reproduces the extracted conductivities without free parameters. This alternate route to the Drude-Smith model provides the popular formula with a more solid physical foundation and well-defined fit parameters.

\end{abstract}

\pacs{Valid PACS appear here}
\maketitle


\section{Introduction}
Terahertz (THz) spectroscopy is a valuable tool for studying charge carrier transport in nanomaterials.\cite{jepsen2010,nemec2010,ulbricht2011,baxter2011,lloyd-hughes2012,joyce2016} THz pulses can act as non-contact, ultrafast probes of intraband excitations in nanoparticle ensembles and provide unique information for next generation nanomaterial engineering. Local nanostructure is revealed via non-Drude THz conductivities, which must be modeled appropriately to gain proper physical insight. While quantum confinement of charge carriers provides an identifiable Lorentzian signature,\cite{ulbricht2011} strong deviations from the Drude model even arise on the mesoscopic scale, where the size of the confining structure is comparable to the carrier mean free path. The term `weak confinement' has been applied to such systems.\cite{ulbricht2011} Interestingly, the appearance of weak-confinement effects in the measured THz conductivity is intrinsically tied to the volume explored by carriers while interacting with the THz pulse. \\ \indent A common estimate for the length scale probed by a THz pulse is the distance a carrier diffuses in one period of the probing frequency, roughly approximated as $L_{\omega} \approx \sqrt{D/\omega}$, where $D$ is the diffusion coefficient and $\omega$ is the angular frequency. Naturally, if structure is present in a sample on a length scale $L_{\omega}$ then it will influence the THz conductivity at a frequency $f=\omega/2\pi$. The challenge is modeling this effect and extracting meaningful physical information.
\\ \indent Over the past decade the Drude-Smith model \cite{smith1968, smith2001} has been highly successful in reproducing the localization signatures observed in a wide variety of materials,\cite{turner2002}$^{-}$\cite{shin2016} including weakly confined systems.\cite{turner2002}$^{-}$\cite{bonetti2016} The Drude-Smith formula is an extension of the phenomenological Drude formula with an additional term that suppresses the conductivity at low frequencies, thereby mimicking the behavior observed in weakly confined systems, where long range carrier transport is suppressed. The Drude-Smith formula is shown in Eq. (1) in the form most commonly employed in the literature,\\
\begin{equation}
\tilde{\sigma} (\omega) = \frac{Ne^{2} \tau_{\text{\tiny{DS}}} / m^{*} }{1-i \omega \tau_{\text{\tiny{DS}}}} \left[ 1 + \frac{c}{1-i \omega \tau_{\text{\tiny{DS}}}}\right] \; \; ,
\end{equation}
where $e$ is the elementary charge, $N$ is the charge carrier density, $m^{*}$ is the carrier effective mass, $\tau_{\text{\tiny{DS}}}$ is the Drude-Smith scattering time (which can differ from the Drude scattering time $\tau$ for a particular material) and $c$ is a constant sometimes referred to as the `localization parameter'. This characterization of $c$ arises because the term in the bracket distinguishes Eq.\,(1) from the bare Drude model formula (and its influence on the conductivity is controlled by the $c$ parameter). $c$ can vary between 0 and -1, where the Drude model is recovered for $c=0$ and the DC conductivity is fully suppressed for $c=-1$. The low-frequency conductivity suppression for $c=-1$ is generally attributed to `carrier backscattering', which has been described both as backwards-biased carrier scattering and as a memory effect, where carriers retain some information of their previous state after scattering, for example because phase coherence is only destroyed after some number of scattering events larger than one.\cite{lovrincic2009} 
\\ \indent The Drude-Smith model in the form shown in Eq. (1) was originally applied to liquid metals \cite{smith1968,smith2001,clerouin2008} and was later adopted for nanosystems,\cite{turner2002}$^{-}$\cite{bonetti2016} where physical boundaries on the nanoscale provide a conceptually clear source of directionally biased scattering. Its use has grown in prevalence over the last few years, extending to disordered crystals,\cite{kang2009,turchinovich2012,iwamoto2013,wang2016} molecular networks,\cite{ai2006,cunningham2008,cunningham2009,unuma2010,cooke2012b,jin2014,yan2016} and high-field transport.\cite{zhou2008,li2012b,shin2016}
\\ \indent Despite its successes, the Drude-Smith model bears two primary criticisms: (i) no rigorous explanation has yet been provided for the assumption that backscattering persists for only one scattering event, which is essential; (ii) the meanings of its fit parameters are unknown beyond phenomenological expressions that depend on multiple physical parameters. Additionally, low-frequency conductivity suppression in some nanomaterials can be explained by alternate theories.\cite{nienhuys2005,strait2009,parkinson2012,boland2016,shimakawa2012,shimakawa2016,mrozek2012,nemec2013,kuzel2014,zajac2014,nemec2015,pushkarev2017}
\\ \indent For example, localized surface plasmon resonances in isolated, photoexcited semiconductor nanowires yield Lorentzian conductivity shapes \cite{nienhuys2005,strait2009,parkinson2012,boland2016} that resemble Drude-Smith conductivities. Still, the two scenarios can be distinguished by the curvature of the experimental complex conductivity and its dependence on carrier density, since this determines the resonance frequency in the plasmonic case ($f_{\circ}\propto\sqrt{N}$). In general, such localized surface plasmon resonances are the result of electrostatic restoring forces. In this scenario, an externally applied electric field, for example the field of a THz probe pulse, displaces charges within a nanoparticle, leading to a `depolarization' dipole field around the nanoparticle that opposes the external field. Joyce \textit{et al}. provide a thorough discussion of THz localized surface plasmon resonances in the context of semiconductor nanowires in Ref. 6.
\\ \indent The localized surface plasmon model accounts for depolarization fields for the specific case of isolated nanostructures of a particular geometry that are composed of a material with a dielectric function that can be described by the Drude model. Conversely, effective medium theories (EMTs) allow for a more general treatment of the local fields around nanostructures. These approaches, such as Maxwell-Garnett or Bruggeman EMT, model the depolarization fields of inhomogeneous media via electrostatics. They incorporate the effects of nanostructure shape, filling fraction, and percolation, as well as the dielectric functions of the materials composing the nanostructure and host matrix. Within the field of THz spectroscopy such EMTs are primarily employed to connect the local fields applied within nanostructures to the measured far-field THz waveform. In principle, this approach makes it possible to extract the average microscopic conductivity of the material within the nanostructures from the measured effective conductivity of the inhomogeneous sample. However, the other degrees of freedom of the system, such as the geometry of the nanostructures and the dielectric function of the host matrix, should be well characterized by independent means, since the use of an EMT significantly increases the number of free parameters.
\\ \indent It is also important to select an appropriate EMT when analyzing a particular sample. Maxwell-Garnett EMT is suitable for evaluating systems where the metallic inclusions in an insulating host matrix are well separated. In fact, as might be expected from their common origin, Maxwell-Garnett EMT reduces to the localized surface plasmon model in the special case where the microscopic conductivity is defined by the Drude model. In this case, the geometry of the metallic inclusions, their filling factor, and the dielectric function of the host matrix contribute to the localized surface plasmon resonance frequency and oscillator strength.\cite{joyce2016} Conversely, Bruggeman EMT is capable of tracking the effective dielectric function of a composite medium across the percolation transition--from a system of isolated metallic inclusions in an insulating host matrix to a system in which long-range carrier pathways exist, and ultimately to a bulk metallic medium. However, neither Bruggeman nor Maxwell-Garnett EMT accounts for weak confinement, and hence this effect must be embedded in the microscopic conductivity of the metallic domains. 
\\ \indent One complication is that it can be difficult to distinguish whether low-frequency conductivity suppression in a particular sample occurs due to depolarization fields, weak confinement, or both. Varying the carrier density can provide some clarity by modifying the depolarization response, in analogy to shifting the resonance frequency in the surface plasmon model. Regardless, it is evident from the literature that weak confinement is an important (and sometimes even dominant) effect in suppressing low-frequency nanomaterial conductivities.\cite{turner2002}$^{-}$\cite{bonetti2016} One clear example of this is a system in which metallic inclusions are present at a density well above the percolation threshold, but in which boundaries also remain, such as a granular film composed of metallic nanoparticles. Such a system will exhibit persistent carrier confinement, resulting in a suppression of the DC conductivity,\cite{cocker2010,titova2016} whereas Bruggeman EMT would predict a Drude-like effective conductivity if the microscopic metallic conductivity were assumed to be Drude.
\\ \indent Since the Drude-Smith model is capable of reproducing the conductivity signature of weak confinement, it has been used as the microscopic conductivity of metallic nanoparticles in EMTs to model THz spectroscopy data.\cite{ahn2007,yang2010,strothkamper2012,zou2012,kim2013,yang2012,nemec2009,nemec2013,zajac2014} Alternatively, the Drude-Smith model has also been used as an EMT in its own right and fit directly to the measured THz conductivity.\cite{turner2002,beard2003,baxter2006,cooke2006,cooke2007,titova2009,titova2011,cooke2012a,walther2007,brandt2008,lovrincic2009,richter2010,cocker2010,cocker2012,tsokkou2012,lim2012,li2012,titova2012,laforge2014,bergren2014,lu2013,buron2014,guglietta2015,yoshioka2015,bonetti2016,titova2016} 
Each approach begins with a different physical process and leads to its own difficulties when one interprets the subsequent fits to the data.
In the former case, depolarization fields are included explicitly and treated as independent from weak confinement. This is problematic because the Drude-Smith formula describes the conductivity of a weakly confined system, i.e. the conductivity of an entire nanoparticle, not just the material inside it. This conductivity includes the influence of boundaries, and as a result changes with nanoparticle size, an effect that can be convoluted with depolarization effects in EMTs. Conversely, in the latter approach, where the Drude-Smith model is applied directly,  it is assumed that depolarization effects are either negligible or can be absorbed into the Drude-Smith fit parameters. This results in significantly fewer fit parameters overall, but also contributes to the general uncertainty surrounding the precise identities of these parameters.
\\ \indent Recent simulations have shown that for a percolated nanoparticle network the macroscopic conductivity is approximately the microscopic conductivity multiplied by a scaling factor,\cite{nemec2013,kuzel2014} indicating that the Drude-Smith model can be applied directly in this case. (Previously, a scaling factor of this type was associated with the metallic filling fraction.\cite{cocker2010}) Nevertheless, depolarization fields do play an important role in determining the macroscopic THz conductivities of some non-percolated systems.\cite{nienhuys2005,strait2009,parkinson2012,boland2016,mrozek2012,nemec2013,kuzel2014,zajac2014,nemec2015} In these cases, the strong dependence of depolarization-based conductivity suppression on carrier density can be used to distinguish it from weak carrier confinement, though it remains unclear how best to combine the two effects. In the following we focus specifically on the microscopic conductivity of weakly confined systems and its relationship to the Drude-Smith model.
\\ \indent It has been demonstrated using Monte Carlo simulations that the microscopic conductivity of weakly confined, classical electrons in the absence of depolarization effects is very similar, though not identical, to the conductivity described by the Drude-Smith formula.\cite{nemec2009,fekete2009} Fitting these simulations enabled Nem\u{e}c \textit{et al}. to provide approximate mathematical expressions for the Drude-Smith fit parameters\cite{nemec2009} that have since been applied to experimental data to extract meaningful physical information.\cite{cocker2010,titova2011,cocker2012,li2012,laforge2014} Complementary experimental techniques are often also used to examine the structure \cite{baxter2006,cooke2006,titova2011,titova2012,titova2016,walther2007,thoman2008,lovrincic2009,ahn2007,richter2010,cocker2010,yang2010,lim2012,li2012,strothkamper2012,yang2012,zou2012,yang2013,laforge2014,kim2013,bergren2014,lu2013,buron2014,yoshioka2015,evers2015,zhang2016,kang2009} and carrier transport characteristics\cite{baxter2006,cooke2006,walther2007,ahn2007,cocker2010,unuma2010,buron2014} in a sample to test the fidelity of the Drude-Smith fits. Nevertheless, in the absence of a tractable derivation that shows how weak carrier confinement yields the Drude-Smith formula with well-understood fit parameters, it is difficult for the Drude-Smith model to be more than a convenient expression for linking the conductivities of similar nanosystems to one another, and to Monte Carlo simulations.
\\ \indent Here, we derive an expression for the microscopic conductivity of a weakly confined Drude gas of electrons and compare its predictions to Monte Carlo simulations of our model system. We find that our analytical conductivity formula agrees with the simulations with no free parameters when the reflectivity \textit{R} of the nanoparticle barriers is \textit{R} = 1. For  0 $<$ $R$ $<$ 1 an ansatz is proposed for the theoretical conductivity, which is found to reproduce the conductivity of the Monte Carlo simulations with one fit parameter. Interestingly, our modified conductivity formula is very similar to the Drude-Smith formula, but contains well-defined parameters. Moreover, we find that the low-frequency conductivity suppression in weakly confined systems is not the result of carrier backscattering off nanoparticle boundaries, but rather arises due to a diffusive restoring current. This current is intrinsically linked to the volume explored by a carrier during one period of the probing frequency and to the build-up of an average local carrier density gradient inside an ensemble of nanoparticles. While this is conceptually different from the conventional interpretation of the Drude-Smith formula, the essential aspect of the derivation was introduced by Smith: a modification to the simple Drude impulse response.\cite{smith1968,smith2001,cooke2012b} We therefore label our new formula as a `modified Drude-Smith' model. 
\\ \indent Our approach in the following is divided into four steps. In the first part, the Monte Carlo simulations are introduced. These differ from the simulations performed by Neme\u{c} \textit{et al}. (Ref. 76) in that we directly track the response of weakly confined electrons to an applied electric field, whereas they used the Kubo formalism. In the second part, we theoretically explore the effects of carrier backscattering for our simulated geometry. In the third part, we present an alternate explanation for the suppression of the low-frequency conductivities in our simulations and derive our modified Drude-Smith model. Finally, we compare the predictions of the modified Drude-Smith model to the conductivities extracted from the Monte Carlo simulations.

\section{Monte Carlo simulations}

Our Monte Carlo simulations are based on the motion of charged, classical, noninteracting particles.  Each particle is described by 
four state variables.  The first and second variables record its position in a two dimensional space, while the third and fourth variables track its velocity in the \textit{x}- and \textit{y}-directions.  At the start of the simulation the particles are initialized with a thermal distribution of velocities.  For a given particle, the components of velocity are assigned independently.  A number is selected from a Gaussian distribution with 0 mean and v$_{\text{th}}^{2}$ variance for each direction, so within our simulation  v$_{\text{th}}\equiv\sqrt{k_{B}T/m^{*}}$.  The ensemble of particles forms a Drude gas of electrons.  In accordance with Brownian motion, the electrons undergo isotropic impurity scattering, corresponding to collisions with phonons and lattice defects.  To model the scattering, a scattering rate $1/\tau$  is introduced, where $\tau$  is the carrier scattering time.  Within our code the probability that a particle will scatter during a particular time step is $\Delta t/\tau$, where $\Delta t$  is the size of a time step in the simulation.  Electron scattering events are independent of one another, as there are no interactions between particles.  If an electron does scatter at a particular time step, it is reinitialized with a thermalized velocity at the \textit{x}- and \textit{y}-position that it occupied when the scattering event occurred.  The \textit{x} and \textit{y} velocity components are selected separately, as was done at the start of the simulation.  A particle therefore has equal probability of scattering in all directions within the \textit{x}-\textit{y} plane.
\\ \indent Weak confinement is introduced via a square box with side length $L$ and reflective boundaries.  At the start of the simulation the particles are initialized at random positions inside the box.  Whenever a particle reaches a boundary, it is either specularly reflected or transmitted through undisturbed.  If the particle is transmitted through the boundary, it appears on the other side of the box, mimicking tunneling or hopping to a new nanoparticle or free motion in the bulk.  The probability of reflection is defined by the parameter \textit{R}, which is an input parameter for the simulation.
\\ \indent To find the conductivity of the system as a function of frequency, the conductivity at each frequency is determined individually using a separate simulation.  An oscillating electric field with frequency $f$ is applied in the \textit{y}-direction and each particle's velocity is recalculated at each time step, where the \textit{y}-component of the velocity at time step n is
\begin{equation}
\text{v}_{\text{y,n}}=\text{v}_{\text{y,n-1}}+\frac{e E_{\text{o}} \Delta t}{m^{*}}\cos \left( 2 \pi f \text{n} \Delta t \right) \; \; ,
\end{equation}
and v$_{\text{y,n-1}}$ is the \textit{y}-component of the velocity at the previous time step, $e$ is the elementary charge, and $E_{\text{o}}$ is the amplitude of the driving field.  A mass of $m^{*}=0.26m_{e}$ is used for the simulations, which is the electron effective mass in silicon.\cite{riffe2002} The velocities of the individual particles are combined to find the net velocity, and hence the net current, at each time step. In the \textit{y}-direction, current oscillations are driven by the electric field, while the current in the \textit{x}-direction provides a noise estimate. The noise is primarily determined by the thermal motion.  For our simulations, we use the realistic estimate of v$_{\text{th}} = 2 \times 10^{5}$\,m/s, where v$_{\text{th}}$ is the one-dimensional thermal velocity v$_{\text{th}}$=$\sqrt{k_{\text{B}}T/m^{*}}$.  The signal-to-noise ratio is affected by the magnitude of the applied electric field through the size of the current response.  For our simulations, we use a peak electric field of 1\,kV/cm, which is in the linear response regime.  
\\ \indent To avoid the transient effects that accompany the turn-on of the electric field, all time steps prior to a total elapsed time of $5 \tau$ are discarded.  A scattering time of 30\,fs is used in our simulations, so $5\tau = 150$\,fs.  The simulation was run for 200,000 time steps for each frequency.  The time step is set to $10^{-16}$\,s, so the total simulation time in each case is 20\,ps.  Hence, at least one period of the driving frequency is contained in every simulation wherein $f \geq 0.05$\,THz.  80,000 particles are used in each simulation.
\begin{figure}  
\begin{center}  
\includegraphics [width=85mm]{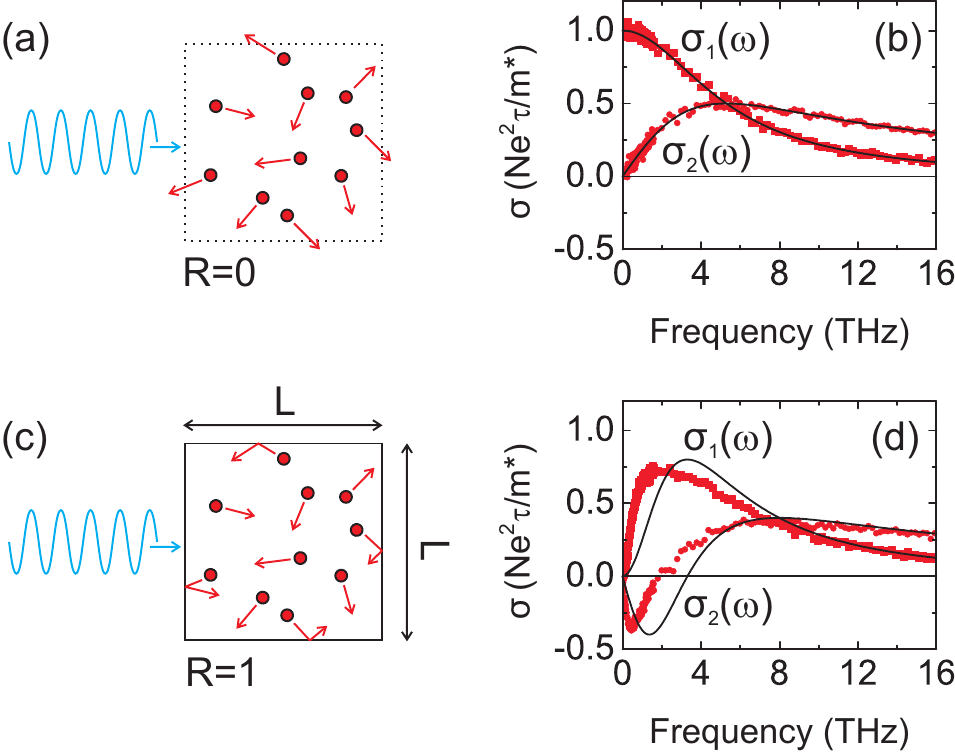}  
\vspace{0.5cm}
\caption{\small \sl Conductivities extracted from the Monte Carlo simulations. Carriers contained in a box with a barrier reflection probability of $R=0$ behave as an unconfined gas of electrons (a) with a Drude conductivity (b).  (c) Increasing the barrier reflectivity to 100\% ($R$=1) results in a suppression of the low-frequency conductivity qualitatively similar to that described by the Drude-Smith model (d). Square points denote the real component of the conductivity and circular points represent the imaginary component of the conductivity. Black lines are fits by the Drude model (b) and Drude-Smith model (d). For both simulations, $\tau$ = 30\,fs, $\text{v}_{\text{th}}$ = 2 $\times$ $10^{5}$\,m/s, $m^{*}$ = $0.26m_{e}$, $L$ = $10\text{v}_{\text{th}}\tau$\,=\,60\,nm, and $E_{\text{o}}$ = 1\,kV/cm. The scattering time of the Drude curve in (b) is 30\,fs, consistent with the simulation. The parameters of the Drude-Smith fit in (d) are $c$\,=\,-1 and $\tau_{\text{\tiny{DS}}}$\,=\,48\,fs. The scattering time $\tau$ in the axes of (b) and (d) refers to the intrinsic scattering time input into the simulations, $\tau$\,=\,30\,fs.\label{fig:Figure 1}}  
\end{center}  
\end{figure}  
\\ \indent Each simulation provides two numbers, the real conductivity ($\sigma_{1}$) and the imaginary conductivity ($\sigma_{2}$) for that particular simulation.  In each simulation, the frequency of the driving field is specified, along with the physical parameters of the electron gas and confining box.  To determine the conductivity for a simulation, we make use of the definition of conductivity $\tilde{\sigma} (\omega) = \tilde{J} (\omega) /\tilde{E} (\omega)$, where $\tilde{\sigma} (\omega)$ is the complex conductivity, $\tilde{J} (\omega)$ is the Fourier transform of the current density, and $\tilde{E} (\omega)$ is the Fourier transform of the electric field.  The simulated current is recorded in the time domain and the value of the electric field is known at all times.  To obtain the conductivity as a function of frequency, the current and electric field are converted to the frequency domain via Fourier transforms,
\begin{equation}
\tilde{\sigma} (\omega) = \frac{ \int^{\infty}_{-\infty} J\left(t\right) e^{i \omega t } dt}{\int^{\infty}_{- \infty} E_{\text{o}} \cos\left(\omega t\right) e^{i \omega t}dt} \; \; .
\end{equation}
Since the simulation is divided into discrete time steps, the integral can be evaluated as a sum.  Making use of the identity $e^{i \omega t}=\cos (\omega t) + i \sin (\omega t)$,
\begin{equation}
\tilde{\sigma} (\omega) = \frac{\sum\limits_{j=1}^{T} J(t_{j}) \left( \cos (\omega t_{j}) + i \sin(\omega t_{j}) \right) }{\sum\limits_{k=1}^{T} E_{\text{o}} \cos(\omega t_{k}) \left( \cos(\omega t_{k}) +i \sin(\omega t_{k}) \right) } \; \; .
\end{equation}
In practice, the sums extend from $j$ = 1500, $k$ = 1500 to $T$ = 200,000, where the starting points of the two indices are increased to account for initial transient effects (over the first $5\tau$).  The fraction in Eq. (4) is broken into real and imaginary parts and summed separately.  Each simulation has a defined driving-field frequency, so Eq. (4) yields a complex number rather than a complex function of frequency. To build up a frequency-dependent conductivity the simulation is repeated for many frequencies. Once the total conductivity has been found for a particular frequency, it is divided by $N e^{2} \tau / m^{*}$, where $\tau$ is the impurity scattering rate that is input as a parameter into the simulation and $N$ is the number of electrons in the simulation.  Based on this normalization, it is expected that when $R$\,=\,0 (Fig. 1(a)) the conductivity will be Drude and the DC conductivity will be 1.  This is indeed the case within uncertainty (defined as the scatter of the conductivity points), as can be seen in Fig. 1(b).  On the other hand, setting the barrier reflectivity to 100\% ($R$\,=\,1, Fig. 1(c)) produces a conductivity qualitatively similar to that described by the Drude-Smith model, as shown in Fig. 1(d), though the Drude-Smith scattering time of $\tau_{\text{\tiny{DS}}}$=48\,fs in Fig. 1(d) does not match the simulation scattering time $\tau$=30\,fs. The connections between the simulation, the Drude-Smith model, carrier backscattering, and carrier diffusion are discussed in detail in the following sections, where we also develop a modified Drude-Smith model that agrees much better with the Monte Carlo simulations over the entire simulation bandwidth, and which requires no free parameters.

\section{Conductivity of a confined Drude gas}

\subsection{Impulse response formalism}

The impulse response formalism provides a convenient framework for our study of weakly confined systems for two reasons. First, we model our confined systems with discontinuous jumps in the potential, and this makes a solution of the underlying differential
equation in the more traditional method more difficult. Second, using the impulse formalism more directly connects our derivation to the works by Smith,\cite{smith1968,smith2001,smith2003} in which he included backscattering to derive the so-called Drude-Smith model. A similar tact was taken by Han \textit{et al.}\cite{han2016} to generalize this model for the magneto-optical conductivity.
\\ \indent The impulse response approach is based on the evolution of current after the application of an impulsive driving force, and can be formulated as follows.  The definition of conductivity is given by
\begin{equation}
\tilde{J}(\omega) = \tilde{\sigma}(\omega)\tilde{E}(\omega) \; \; ,
\end{equation}
where the complex notation allows us to account for the current response in and out of phase with the driving electric field.  By the convolution theorem, Eq. (5) can be rewritten as 
\begin{equation}
J(t)=\int_{-\infty}^{\infty}\sigma(t-t') E(t') dt' \; \; .
\end{equation}
Within the impulse response formalism, an impulse of electric field is applied to the system such that 
\begin{equation}
E(t')=E_{\text{i}} \delta(t') \; \; ,
\end{equation}
where it should be noted that the quantity $E_{\text{i}}$ is the magnitude of the impulse and has units of electric field $\times$ time, in accordance with the units of the time-dependent electric field $E(t)$ and delta function $\delta (t)$.  Substituting the electric field impulse into Eq. (6), 
\begin{equation}
 J(t) = \int_{-\infty}^{\infty} \sigma(t-t') E_{\text{i}} \delta(t')dt' = \sigma(t) E_{\text{i}} \; \; .
\end{equation}
The impulse-response function is thus defined as
\begin{equation}
\frac{J(t)}{E_{\text{i}}} = \sigma(t) \equiv j(t) \; \; .
\end{equation}
Furthermore, determining the complex conductivity $\tilde{\sigma} (\omega)$ from $j(t)$ is straightforward, as
 \begin{equation}
\tilde{\sigma}(\omega) = \int^{\infty}_{-\infty} j(t) \exp(i \omega t) dt \; \; .
\end{equation}
Therefore, provided the impulse-response function $j(t)$ (hereafter referred to simply as the impulse response) can be determined, the conductivity of the system can be found via the Fourier transform of $j(t)$.  Furthermore, the magnitude of the impulse response at $t$ = 0 is given by\cite{smith2001}
\begin{equation}
j(0) = \frac{J(0)}{E_{\text{i}}} = \frac{Ne^{2}}{m^{*}} \; \; ,
\end{equation}
where $N$ is the electron density. An added benefit of the impulse response approach is that the Kramers-Kronig relations are automatically satisfied because causality is introduced at the outset.\cite{smith2001} Finally, we note that $j(t)$ is essentially the time-dependent current density induced by the impulse, though lower case notation is used to maintain proper units.

\subsection{Drude-Smith model}

The Drude model can be easily obtained via the impulse response approach by considering isotropic scattering: for every carrier that undergoes a collision, be it with a phonon, a lattice impurity or another electron, resulting in a new velocity in the forward direction, another carrier will undergo a collision that results in a velocity in the backward direction. Consequently, the contribution to the impulse response from particles that have scattered will be on average zero. Since only unscattered particles contribute to the impulse response, scattering acts to decrease the population of current-carrying particles.  If the carrier scattering rate is $1/\tau$, then the resulting impulse response is
\begin{equation}
j(t)/j(0) = \exp(-t/\tau) \Theta (t) \; \; ,
\end{equation}
where $\Theta (t)$ is the unit step function, which ensures that no current is present prior to the impulse.  Mathematically, its role is to shift the limits of integration in the Fourier transform connecting $j(t)$ to $\tilde{\sigma}(\omega)$  from $\left( -\infty , \infty\right)$ to $\left[ 0,\infty \right)$.  Taking the Fourier transform of Eq. (12) gives the well-known Drude conductivity formula given by
\begin{equation}
\tilde{\sigma}(\omega) = \frac{Ne^2 \tau /m^{*}}{1-i\omega \tau} \; \; .
\end{equation}
\indent In the Drude-Smith model carrier scattering is not isotropic, and hence scattering cannot be treated as a simple population decay in the impulse response formalism. Rather, the probability $p$ that a carrier has undergone n scattering events in the time interval from 0 to $t$ is modeled using Poisson statistics, such that
\begin{equation}
p_{\text{n}}(0,t)=(t/\tau)^{\text{n}} \exp(-t/ \tau)/\text{n}! \; \; ,
\end{equation}
for a scattering rate 1/$\tau$. To describe biased carrier backscattering, Smith introduced the set of parameters $c_{\text{n}}$. The index n represents the scattering event, where n\,=\,1 is the first scattering event for a given carrier, n\,=\,2 is the second scattering event, and so on.  Each $c_{\text{n}}$ can be viewed either as a measure of the memory a carrier sustains of its previous velocity, or alternatively, as the probability that it will backscatter ($\times (-1)$) rather than scatter isotropically.  The $c_{\text{n}}$ parameters can range from $-1$ to 0, where complete backscattering occurs for $c_{\text{n}}=-1$ and isotropic scattering is recovered for $c_{\text{n}}=0$.  The impulse response for the Drude-Smith model is
\begin{equation}
\frac{j(t)}{j(0)}= \exp (-t/ \tau ) \Theta(t)  \times \left[ 1 + \sum \limits_{\text{n}=1}^{\infty} c_{\text{n}} (t/ \tau)^{\text{n}}/\text{n}! \right] .
\end{equation}
If $c_{\text{n}}=-1$ then the $\text{n}^{th}$ collision of a particle will result in it scattering back in the direction it came from.  Consequently, the contribution of said particle to the average current will change sign.  Taking the Fourier transform of Eq. (15) gives
\begin{equation}
\tilde{\sigma} (\omega) = \frac{Ne^{2} \tau / m^{*} }{1-i \omega \tau} \left[ 1 + \sum \limits_{\text{n}=1}^{\infty} \frac{c_{\text{n}}}{(1-i \omega \tau)^{\text{n}}}\right] \; \; ,
\end{equation}
which is the general form of the Drude-Smith conductivity formula.  However, Eq. (16) is not the conductivity most commonly associated with the Drude-Smith model.  To obtain the familiar, truncated Drude-Smith conductivity formula, Smith made a key assumption, namely that the backscattering bias exists only for the first scattering event.  Under this approximation, $c_{\text{n}>1}=0$ and $c_{1}$ is relabelled as $c$.  The resulting series truncation yields the Drude-Smith impulse response,
\begin{equation}
j(t)/j(0) = \exp (-t/ \tau ) \left[ 1 +  \frac{ct}{\tau}  \right]\Theta (t) \; \; .
\end{equation}
The truncated Drude-Smith conductivity formula can be found via the Fourier transform of this impulse response,
\begin{equation}
\tilde{\sigma} (\omega) = \frac{Ne^{2} \tau / m^{*} }{1-i \omega \tau} \left[ 1 + \frac{c}{1-i \omega \tau}\right] \; \; ,
\end{equation}
which is the formula commonly used for fitting experimental conductivities\cite{turner2002}$^{-}$\cite{shin2016} that was introduced in Section\,I (Eq.\,(1) with the free fit parameter $\tau_{\text{\tiny{DS}}}$ set to $\tau$).
\\ \indent However, Smith also noted that if the scattering events are independent of one another, the single-scattering approximation is not valid.\cite{smith2001}  Under the alternate interpretation that the scattering events are all equivalent, $c_{\text{n}}=c^{\text{n}}$ and the infinite series in the impulse response (Eq. (15)) can be found exactly,
\begin{align}
j(t)/j(0) &= \exp(-t/ \tau) \sum \limits_{\text{n}=0}^{\infty} \frac{\left( ct / \tau \right)^{\text{n}}}{\text{n}!}\Theta (t) \nonumber \\ \nonumber \\ &= \exp(-t/ \tau) \exp(ct/ \tau) \Theta (t) \nonumber \\ \nonumber \\ &=\exp(-t/\tau ') \Theta (t) \; \; ,
\end{align}
where $\tau' = \tau/(1-c)$.  Since evaluating the sum in the impulse response yields an exponential decay, the Drude-Smith model collapses to the Drude model with a modified scattering time,
\begin{equation}
\tilde{\sigma}(\omega) = \frac{Ne^2 \tau ' /m^{*}}{1-i\omega \tau '} \; \; .
\end{equation}
\\ \indent Reliance on the single backscattering approximation has led to criticisms of the physicality of the truncated Drude-Smith formula.\cite{nienhuys2005,shimakawa2012,nemec2013,kuzel2014,zajac2014} However, we note that the approach introduced by Smith applies generally to any differentiable function that modifies the Drude impulse response, and is not specific to the case of carrier backscattering.\cite{smith2001,cooke2012b} It is possible, therefore, that the truncated Drude-Smith conductivity shape observed in real systems arises due to a somewhat different physical mechanism than originally proposed. This is explored further in the following sections. Here, we predict the conductivity of the weakly confined Drude gas of electrons in our Monte Carlo simulations based on the above derivation of the Drude-Smith model. To describe our Monte Carlo simulations, two separate scattering mechanisms should be included: bulk, isotropic impurity scattering and specular boundary scattering. Impurity scattering is taken to occur at approximately the same rate as in bulk, though this is not a necessary condition. Boundary scattering, on the other hand, occurs at the rate that carriers encounter the nanoparticle barriers, and in contrast to impurity scattering, the probability that a given carrier has undergone n boundary scattering events in the time interval from 0 to $t$ is unequivocally not given by Poisson statistics. However, as a first approximation, we incorporate boundary scattering as a secondary scattering event with a backscattering bias in the Poisson-statistics-based Drude-Smith impulse response, such that 
\begin{align}
\frac{j(t)}{j(0)}&= \exp (-t/ \tau ) \exp (-t/ \tau_{\text{B}}) \\& \nonumber \times \left[ 1 + \sum \limits_{\text{n}=1}^{\infty} c_{\text{n}} (t/ \tau_{\text{B}})^{\text{n}}/\text{n}! \right]\Theta (t) \; \; ,
\end{align}
where 1/$\tau_{\text{B}}$ is the boundary scattering rate. We note that the validity of employing Poisson statistics to describe boundary scattering in our simulations is explored in the following section. Within the current approach, scattering will always occur when a carrier reaches a nanoparticle boundary.  The $c_{\text{n}}$ parameters bias the scattering events between isotropic ($c_{\text{n}}=0$) and reflective ($c_{\text{n}}=-1$).  Furthermore, because interactions with the boundaries are equivalent within our simple model system, $c_{\text{n}}=c^{\text{n}}$.  While it is possible that some unforeseen detail could break this symmetry, we expect that the Drude-Smith conductivity should reduce to Eq.\,(20) with an effective scattering time of
\begin{equation}
\tau ' = \left( \frac{1}{\tau} + \frac{1-c}{\tau_B} \right) ^{-1} \; \; .
\end{equation}
\\ \indent In actuality, the boundaries in our Monte Carlo simulations are either reflective or transmissive, and not scattering centers with a potential direction bias. We therefore refine our treatment of nanoparticle barrier scattering: carriers that encounter a boundary either rebound off it or pass through it unaffected, with a probability defined by the reflection coefficient of the boundary. Within our Monte Carlo simulations, only confinement in the $y$-direction affects the conductivity. The electric field is oriented along the \textit{y}-direction, so the top and bottom walls of the box are oriented perpendicular to the electric field. Carriers that bounce off these walls--an event which occurs with a probability $R$--undergo collisions wherein $c_{\text{n}} = (-1)^{\text{n}}$. The reflection coefficient ($R$), on the other hand, changes the boundary scattering rate, so the average collision time with a boundary is given by $L/(R \text{v}_{\text{th}})$.  Recall that $\text{v}_{\text{th}}$ is defined as the root mean square speed in a single direction within our Monte Carlo simulation and  $L$ is the width of the box.  Therefore, the Drude-Smith impulse response for our simulated system should be
\begin{align}
\frac{j(t)}{j(0)}&= \exp (-t/ \tau ) \exp (-R\text{v}_{\text{th}}t/ L) \\& \nonumber \times \left[ 1 + \sum \limits_{\text{n}=1}^{\infty} (-R\text{v}_{\text{th}}t/ L)^{\text{n}}/\text{n}! \right] \Theta (t) \; \; ,
\end{align}
and since the sum is again equivalent to an exponential decay, the conductivity of our system predicted by the full Drude-Smith model formalism is Drude (Eq. (20)) with an effective scattering time given by
\begin{equation}
\tau '= \left( \frac{1}{\tau} + \frac{2R\text{v}_{\text{th}}}{L} \right) ^{-1} \; \; .
\end{equation}
\indent While the conductivity predicted by the above formalism for our model system seems obvious (Eq. (20)), this prediction is clearly incorrect. The example conductivity shown in Fig. 1(d) for a Monte Carlo simulation with $R=1$ is drastically non-Drude-like, and in fact resembles the shape of the truncated Drude-Smith model for $c=-1$. The strongest argument for the series truncation is therefore that the resulting Drude-Smith conductivity resembles the simulated conductivity, and this is also often the case for experimental fitting. However, we also note that the fit parameter $\tau_{\text{\tiny{DS}}}$ in Fig\,1(d) does not agree with the simulation parameters. Below, we deduce the precise effect of carrier backscattering on the conductivity of our model system to determine whether the series truncation can be justified in this case.

\subsection{Carrier backscattering}

Here, we reassess the effect of carrier backscattering on the conductivity of the model system described in Section II. In particular, the impulse response is evaluated explicitly for the geometry of the Monte Carlo simulations. One purpose of this exercise is to determine whether boundary scattering is accurately described by Poisson statistics or if some justification for the Drude-Smith series truncation is introduced by the regularity of the boundary scattering interval. Structural effects are incorporated directly into the impulse response through an evaluation of the current evolution in the time domain.
\\ \indent Our thought experiment begins with a single, classical charged particle sitting in the center of a one-dimensional box of width $L$ with boundary reflection probability $R$\,=\,1.  If an impulse to the right is provided to the particle, the resulting impulse response will be a square wave with a period of $2L/\text{v}(0)=2Lm^{*}/eE_{\text{i}}$, corresponding to the round-trip time of the particle, where $\text{v}(0)=J(0)/(Ne)$ is the speed imparted to the particle.  The impulse response is shown in Fig. 2(a).
\\ \indent If the box is alternatively filled with charged, non-interacting particles at rest, an impulse will result in a triangle-wave impulse response with a period of $2Lm^{*}/eE_{\text{i}}$.  (A box filled with non-interacting particles is equivalent to many boxes, each containing a single particle initialized at a random position.)  The impulse response can be proven to be a triangle wave by simple geometry arguments, or via an integration over phase of square waves with infinitesimal amplitude.  The triangle wave impulse response is shown in Fig. 2(b).
\begin{figure}  
\begin{center}  
\includegraphics [width=85mm]{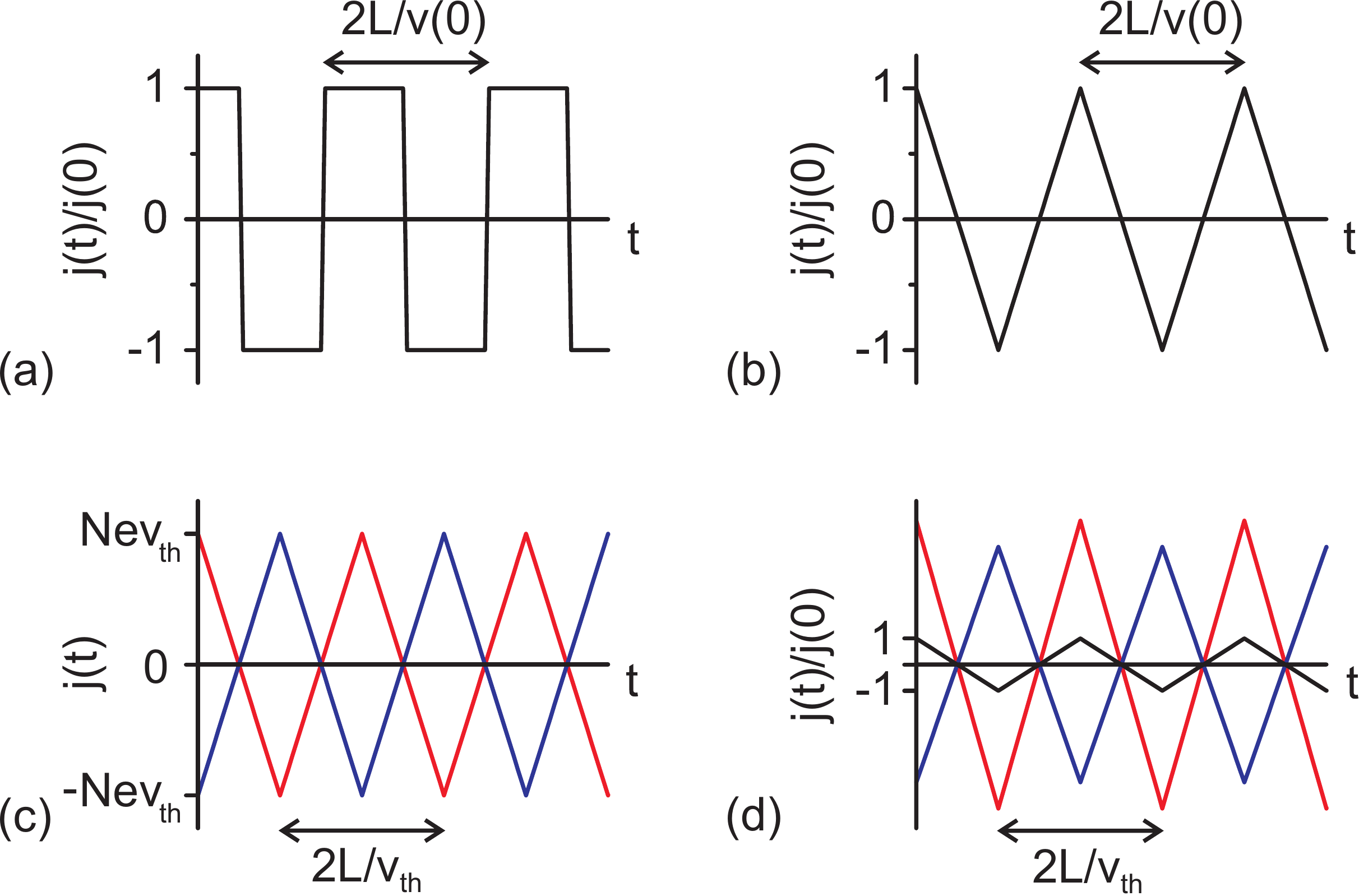}  
\vspace{0.5cm}
\caption{\small \sl Ballistic impulse responses of classical particles.  (a)\,An impulse applied to a single, stationary particle at the center of a one-dimensional box of width $L$ with boundary reflection probability $R$\,=\,1 will produce a square-wave impulse response.  (b) The impulse response for a box filled with noninteracting particles initially at rest will be a triangle wave.  (c) A box filled with particles that have either v\,=\,v$_{\text{th}}$ or v\,=\,-v$_{\text{th}}$ will have zero net current because the initially right-moving and initially left-moving particles will trace out offsetting triangle wave currents (blue and red curves).  (d) An impulse applied to the system in (c) will result in a triangle-wave impulse response with a period of approximately $2L/\text{v}_{\text{th}}$ (black curve). \label{fig:Figure 2}}  
\end{center}  
\end{figure} 
\\ \indent The picture of initially stationary particles does not accurately reflect the physics of our model system. Thus, thermal motion is introduced.  As before, the one-dimensional box is filled with non-interacting particles, but they are now initialized with a background velocity.  Half the particles travel right at a speed v$_{\text{th}}$ and half the particles travel left at {v}$_{\text{th}}$.  Only a single speed is present, as a thermal distribution has not yet been introduced.  Prior to the application of the impulse, the background current is zero, as shown in Fig. 2(c).  The set of particles initially traveling right produces a triangle-wave current with a period $2L/\text{v}_{\text{th}}$ (red curve in Fig. 2(c)).  The set of particles initially traveling left produces a triangle wave current with identical amplitude and period, but with a $180^{\circ}$ phase offset (blue curve in Fig. 2(c)).  Therefore, the net current is zero (black horizontal line in Fig. 2(c)). 
\\ \indent Upon the application of an impulse to the right at $t=0$ a net current is produced, as shown in Fig. 2(d).  The total current from the particles initially moving right will again be a triangle wave, but with an amplitude increased by the net current impulse provided to the right-moving particles.  The period of the triangle wave is $2L/\text{v}_{\text{th}}-2Lm^{*}/eE_{\text{i}}$.  The current due to the initially right-moving particles is depicted by the red curve in Fig.\,2(d).  The blue curve in Fig.\,2(d) corresponds to the current from the initially left-moving particles, which is a triangle wave with an amplitude reduced by the net current impulse provided to the left-moving particles and a period of $2L/\text{v}_{\text{th}}+2Lm^{*}/eE_{\text{i}}$.  The net impulse response is the sum of the two triangle waves. 
\\ \indent If $\text{v}_{\text{th}}\gg eE_{\text{i}}/m^{*}$, that is if the thermal speed is much larger than the speed provided to a particle by the impulse, then the two triangle waves have approximately the same period.  This approximation is physically reasonable based on the characteristic free carrier velocities in semiconductors and metals and the peak electric field of a typical THz pulse used for spectroscopy.  For instance, a THz pulse with a common peak field of 1\,kV/cm will produce a drift current in silicon on the order of $10^{3}$ m/s for a carrier scattering time of 30 fs, compared to a carrier thermal velocity on the order of $10^{5}$\,m/s.  Therefore the impulse response is approximately a triangle wave with a period of $2L/\text{v}_{\text{th}}$ and an amplitude of $j(0)$.  This triangle-wave impulse response is shown in black in Fig. 2(d).  Defining  $t_{\circ} \equiv L/\text{v}_{\text{th}}$, the impulse response can be written as
\begin{align}
\frac{j_{\circ}(t)}{j(0)} &=  \sum^{\infty}_{\text{n}=0}  \Theta(t-\text{n}t_{\circ})\Theta((\text{n}+1)t_{\circ}-t)  \\ & \nonumber \times \left[(-1)^{\text{n}}\left( \frac{-2}{t_{\circ}}t+(1+2\text{n})\right) \right] \; \; .
\end{align}
The impulse response is labeled $j_{\circ}(t)$ here for notation reasons that will be made clear in the following section.
\\ \indent Equation (25) describes the response of a ballistic system to an impulse.  However, carriers contained within a nanoparticle will experience regular impurity scattering (off lattice defects and phonons) in addition to nanostructure boundary scattering.  In bulk, these scattering events are isotropic, as discussed in the derivation of the Drude model using the impulse response formalism (Eqs.\,(12) and (13)). When nanostructure is introduced, though, the situation changes. Impurity scattering that is intrinsically isotropic becomes effectively anisotropic near nanostructure boundaries because particles scattered towards a boundary are reflected back in the opposite direction. Therefore, accounting for impurity scattering in a nanoparticle with an exponential decay in the impulse response actually models particles that freeze in place where they scatter, rather than particles that scatter isotropically like in the Drude model. As a first approximation, however, we describe impurity scattering in our model system in this way, multiplying the ballistic impulse response by an exponential decay, 
\begin{align}
\frac{j_{\circ}(t)}{j(0)}&=\sum^{\infty}_{\text{n}=0} \Theta(t-\text{n}t_{\circ})\Theta((\text{n}+1)t_{\circ}-t)  \\ & \nonumber \times \left[ (-1)^{\text{n}} \left( \frac{-2}{t_{\circ}}t +(1+2\text{n})\right) \right] e^{-t/\tau} \; \; .
\end{align}
where $\tau$ is the impurity scattering time.  In the following section, we improve on this approximation by introducing diffusion, which allows the particle density to return to its equilibrium distribution. 
\\ \indent The conductivities corresponding to the impulse responses in Eq. (25) and Eq. (26) are obtained by taking their Fourier transforms. Whereas the conductivity for the ballistic case can be found using the Fourier series of an infinite triangle wave (and corresponds to an infinite series of Dirac delta functions, each with an associated divergence in the imaginary conductivity) impurity scattering makes the impulse response in Eq. (26) nonperiodic, so the Fourier transform must be found directly, i.e.
\begin{align}
\tilde{\sigma}_{\circ}(\omega) &= \frac{Ne^{2}}{m^{*}} \sum^{\infty}_{\text{n}=0} (-1)^\text{n}  \\ & \nonumber \times \int^{(\text{n}+1)t_{\circ}}_{\text{n}t_{\circ}} dt \left[  \frac{-2}{t_{\circ}} t +(1+2\text{n})\right] e^{(i\omega -1/\tau)t} \; \; .
\end{align}
 The integral in Eq. (27) can be evaluated exactly, giving a geometric series that converges (only) for finite $\tau$ to
\begin{align}
\tilde{\sigma}_{\circ}(\omega) &= \frac{Ne^{2}}{m^{*}} \frac{\tau}{1-i\omega\tau}  \\ & \nonumber \times \left[ 1+\frac{2\tau /t_{\circ}}{1-i\omega \tau} \left( \frac{e^{-t_{\circ}/\tau}e^{i\omega t_{\circ}} -1}{e^{-t_{\circ}/\tau}e^{i\omega t_{\circ}} +1}\right) \right] \; \; .
\end{align}
Equation (28) has been derived based on 100\% carrier reflecting barriers, but in our Monte Carlo simulation there is a finite probability of carrier transmission through the barriers if $R\neq 1$.  This mimics carrier hopping to neighboring nanoparticles, and is important because many nanomaterials with weak confinement will exhibit some degree of long range, DC transport, albeit suppressed.
\\ \indent To introduce the added complexity of partially reflecting barriers, the net ballistic current that persists in the direction of the impulse is examined as a function of time.  If the n$^{th}$ encounter with the boundaries is defined as the moment the last particle in the box undergoes its n$^{th}$ collision, then $j_{\circ}(\text{n})/j(0)=(1-2R)^{\text{n}} $.  This relation was determined recursively, where the total current was calculated for a series of boundary interactions until a pattern emerged and was confirmed.  Hence, the generalized impulse response for barriers with a reflection probability of $R$ is
\begin{align}
\frac{j_{\circ}(t)}{j(0)} &= \sum^{\infty}_{\text{n}=0} \Theta (t-\text{n}t_{\circ}) \Theta((\text{n}+1)t_{\circ}-t) \\ & \nonumber \times  \left[(1-2R)^{\text{n}} \left( \frac{-2R}{t_{\circ}} t +(1-2\text{n}R)\right)\right] e^{-t/\tau} \; \; .
\end{align}
The Fourier transform of the impulse response can be evaluated exactly, as in the $R=1$ case, giving
\begin{align}
\tilde{\sigma}_{\circ}(\omega) &= \frac{Ne^{2}}{m^{*}} \frac{\tau}{1-i\omega\tau} \\ & \nonumber \times  \left[ 1+\frac{2\tau R /t_{\circ}}{1-i\omega \tau} \left( \frac{e^{-t_{\circ}/\tau}e^{i\omega t_{\circ}} -1}{1-(1-2R) e^{-t_{\circ}/\tau}e^{i\omega t_{\circ}}}\right) \right] \; \; .
\end{align}
\\ \indent The extension of our model system to multiple dimensions and the generalization to a thermal distribution of velocities can be accomplished in a single step.  In our Monte Carlo simulation the boundaries are parallel to the $x$- and $y$-directions, and hence boundary scattering preserves the speed in each direction. Since the applied electric field is parallel to the $y$-direction, confinement only matters in this direction. Furthermore, the equipartition theorem ensures that the two particle velocity components are independent, and therefore a thermal distribution is required only in the $y$-direction. Furthermore, note that v$_{\text{th}}$ is defined to be $\sqrt{k_{B}T/m^{*}}$, independent of dimension.
\\ \indent To find a generalized impulse response, the impulse response for a single speed in the $y$-direction is averaged over all possible $y$-direction speeds, weighted by a thermal distribution.  We introduce the variable velocity in the $y$-direction $\text{v}_{y}$, which results in a transit time across the box of $L/\text{v}_{y}$, and note that v$_{\text{th}}$ is the variance of the thermal weighting.  The integration over velocities is limited to the range $\left[ 0,\infty\right)$  because a distinction was made between the forward and backward thermal velocities in the derivation of $j_{\circ}$.  Thus,
\begin{equation}
\langle j_{\circ} (t) \rangle = \sqrt{2 \over \pi \text{v}^2_{\text{th}}} \int_{0}^{\infty} e^{-\text{v}_{y}^{2}/2\text{v}_{\text{th}}^2}j_{o}(\text{v}_{y},t)d\text{v}_{y} \; \; .
\end{equation}
As usual, to obtain the conductivity of the model system from the impulse response we take the Fourier transform,
\begin{equation}
\langle \tilde{\sigma}_{\circ}(\omega) \rangle = \sqrt{2 \over \pi \text{v}^2_{\text{th}}} \int_{-\infty}^{\infty} e^{i \omega t} dt \int_{0}^{\infty}  e^{-\text{v}_{y}^{2}/2\text{v}_{\text{th}}^{2}} d\text{v}_{y} \left[ j_{\circ}(\text{v}_{y},t) \right]  .
\end{equation}
The order of the two integrals is interchangeable, so Eq.\,(32) can be rewritten as
\begin{equation}
\langle \tilde{\sigma}_{\circ}(\omega) \rangle = \sqrt{2 \over \pi \text{v}^2_{\text{th}}}\int_{0}^{\infty}  e^{-\text{v}_{y}^{2}/2\text{v}_{\text{th}}^{2}} d\text{v}_{y} \left[ \tilde{\sigma}_{\circ}(\text{v}_{y},\omega) \right] \; \; .
\end{equation}
The conductivity formula written in detail is 
\begin{align}
&\langle \tilde{\sigma}_{o}(\omega) \rangle =  \frac{Ne^{2}}{m^{*}} \frac{\tau}{1-i\omega\tau} \sqrt{2 \over \pi \text{v}^2_{\text{th}}} \int_{0}^{\infty}  e^{-\text{v}_{y}^{2}/2\text{v}_{\text{th}}^{2}} d\text{v}_{y} \\ \nonumber & \times \left[ 1+\frac{2\tau R \text{v}_y/L}{1-i\omega \tau} \left( \frac{e^{-L/\text{v}_{y}\tau}e^{i\omega L/\text{v}_{y}} -1}{1-(1-2R) e^{-L/\text{v}_{y}\tau}e^{i\omega L/\text{v}_{y}}}\right) \right] \; \; .
\end{align}
Note that for simplicity the integration over the magnitude of the velocity starts at $\text{v}_y = 0$, even though we are assuming  a weak electric field. We judge this to be physically reasonable because deviations from $\text{v}_y >> eE_{\text{i}}/m^{*}$ only occur in the low energy tail of the thermal distribution.
\\ \indent We have directly considered the effects of boundary scattering on the impulse response to obtain Eq. (34) and in doing so we have removed the uncertainty associated with the role of boundary backscattering.  This should provide a critical test of whether the collapsed or truncated Drude-Smith formula correctly describes backscattering in our model system of weakly confined electrons.
\\ \indent Equation (34) can be evaluated numerically provided the parameters $Ne^{2}/m^{*}, \tau, R, L,$ and v$_{\text{th}}$ are specified.  Therefore, the results of our alternate derivation can be compared to the predictions based on the Drude-Smith formalism.  From a mathematical standpoint the collapsed Drude-Smith model is expected (Eqs. (20) and (24)), while the interpretation of the Drude-Smith model often used in experimental fitting requires that the truncated form emerges (Eqs. (1) and (18)).  We stress that no fit parameters are needed for the comparison, as all the variables are provided.

\begin{figure}  
\begin{center}  
\includegraphics [width=85mm]{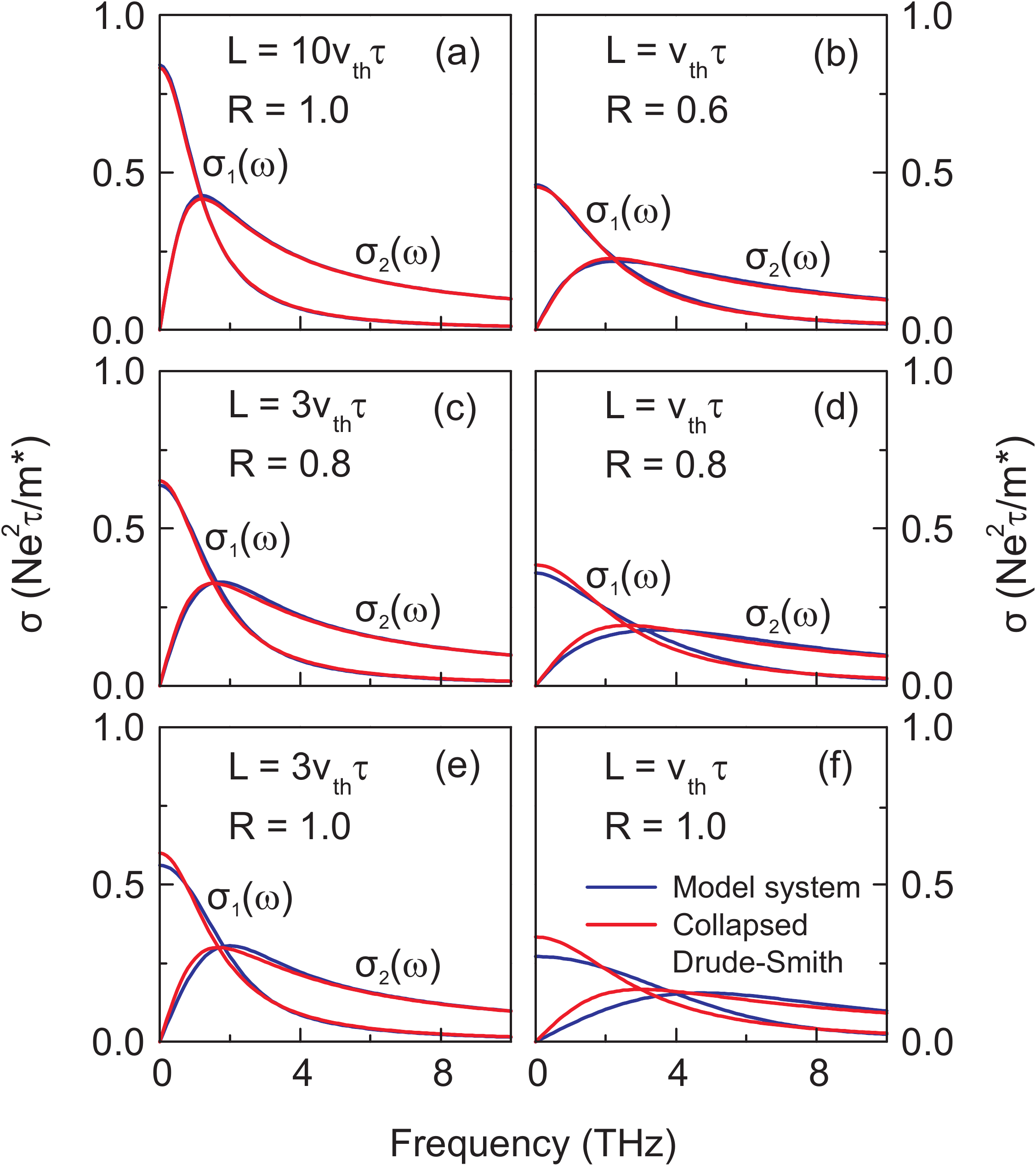}  
\vspace{0.5cm}
\caption{\small \sl Conductivity of a confined Drude gas of electrons including carrier backscattering off boundaries for a selection of box sizes (L) and barrier reflectivities (R). Blue curves show the complex conductivity ($\sigma_{1}$ and $\sigma_{2}$) of the model system given by Eq. (34). Red curves show the complex conductivity predicted by the Drude-Smith model, where all boundary scattering events are taken to be equivalent and the formula collapses to the Drude model (Eq. (20)) with a modified scattering time (Eq. (24)). $\tau=30$\,fs; v$_{\text{th}}=2 \times 10^{5}$\,m/s. \label{fig:Figure 3}}  
\end{center}  
\end{figure} 

As can be seen in Fig.\,3(a)-(f), the conductivities predicted for a confined Drude gas of electrons by the collapsed Drude-Smith model (Eq. (20)) and Eq. (34) agree over a large parameter space.  Figure 3 shows the results of the two approaches for box widths of $L=\text{v}_{\text{th}}\tau$\,=\,6\,nm, $L=3\text{v}_{\text{th}}\tau$\,=\,18\,nm, and $L=10\text{v}_{\text{th}}\tau$\,=\,60\,nm, and barrier reflectivities ranging from $R=0.6$ to $R=1$. For $R<0.6$ the two conductivity curves are nearly identical for $L \geq \text{v}_{\text{th}}\tau$ (not shown). The only parameter set that results in significant discrepancy between the two theories is $L=\text{v}_{\text{th}}\tau$, $R=1$, as shown in Fig. 3(f).  While the difference between the curves is not large, it may indicate that for small boxes with highly reflective barriers the number of boundary scattering events experienced by a given particle is not well described by Poisson statistics.
\\ \indent For all sets of parameters tested we recover the Drude model for the conductivity of a weakly confined system based on backscattering physics, suggesting that the truncation of the series in the derivation of the Drude-Smith model is not justified for this system. Nevertheless, the conductivity of the Monte Carlo simulations for $R\neq 0$ is non-Drude and exhibits prominent conductivity suppression at low frequencies as $R$ approaches 1. The physics that is still missing is revealed by Fig. 3: the DC conductivity of electrons contained in a box with 100\% reflecting barriers must be zero since long-range transport is impossible if a barrier is present that prevents all particles from passing through, but this is not the case for either Eq. (20) or Eq. (34) (see Figs. 3(a), (e), and (f)). Clearly, there is an important aspect to the system that has not yet been included.  
\\ \indent The presence of a DC conductivity implies a shift in average particle position due to an impulse.  For our model system, an initial shift due to the impulse is expected, but afterwards the average particle position will be restored to the center of the box. The mechanism that restores the average particle position, i.e., diffusion, is exactly the physics that has been missing from the model to this point. In the next section, we show how diffusion acts to restore the average particle position after an impulse or to counteract the drift current induced by a continuous electric field, producing the low-frequency suppression seen in weakly confined systems.

\subsection{Diffusion restoring current}

Consider the case of weak carrier confinement in a box with 100\% reflecting boundaries.  The shift in average particle position induced by an impulse will result in a change to the carrier density profile within the box because the carriers are restricted from moving beyond its walls.  If a constant DC electric field were applied rather than an impulse, the carrier density profile would be exponential as a function of position in the $y$-direction.  However, for a sufficiently small electric field ($\text{v}_{\text{drift}}/\text{v}_{\text{th}} \ll 1$) the density profile can be approximated as linear, and we assume that this is also the case for the carrier density profile that forms following an impulse.
\\ \indent Figure 4 illustrates the carrier density $N$ as a function of position in the $y$-direction, parallel to the applied field in our Monte Carlo simulations.  For a linear density profile, the carrier density as a function of $y$-position is given by
\begin{equation}
N=N_{\circ} + y\frac{dN}{dy} \; \; ,
\end{equation}
where $N_{\circ}$ is the average electron density in the box, and is equal to the electron density in the undisturbed system.  The average particle position is
\begin{equation}
\overline{y} = \frac{L^{2}}{12N_{\circ}}\frac{dN}{dy} \; \; ,
\end{equation}
where the center of the box is defined as the origin.  The shift in average particle position due to an impulse can be written in terms of the impulse response, since the average particle velocity is given by $\text{v}(t)=j(t)E_{\text{i}}/N_{\circ}e$.  Thus,
\begin{equation}
\overline{y} = \frac{E_{\text{i}}}{N_{\circ}e} \int^{t}_{0} j(t') dt' \; \; .
\end{equation}
In general, if a carrier density gradient exists, a current due to diffusion will also be present.  Diffusion requires no electron-electron interactions or electrostatic fields; it is a consequence of probability.  The diffusion current is proportional to particle density gradient, such that
\begin{equation}
J_{\text{diff}} = -eD \frac{dN}{dy} \; \; ,
\end{equation}
where $D$ is the diffusion coefficient.  Since the impulse induces a carrier density gradient, a diffusion component will in turn be present within the impulse response, 
\begin{equation}
j_{\text{diff}}(t) = \frac{-12D}{L^{2}} \int_{0}^{t} j(t') dt' \; \; ,
\end{equation}
which is obtained by combining Eqs. (36), (37), and (38), and defining $j_{\text{diff}}(t) \equiv J_{\text{diff}}$/ $E_{\text{i}}$.

\begin{figure}  
\begin{center}  
\includegraphics [width=55mm]{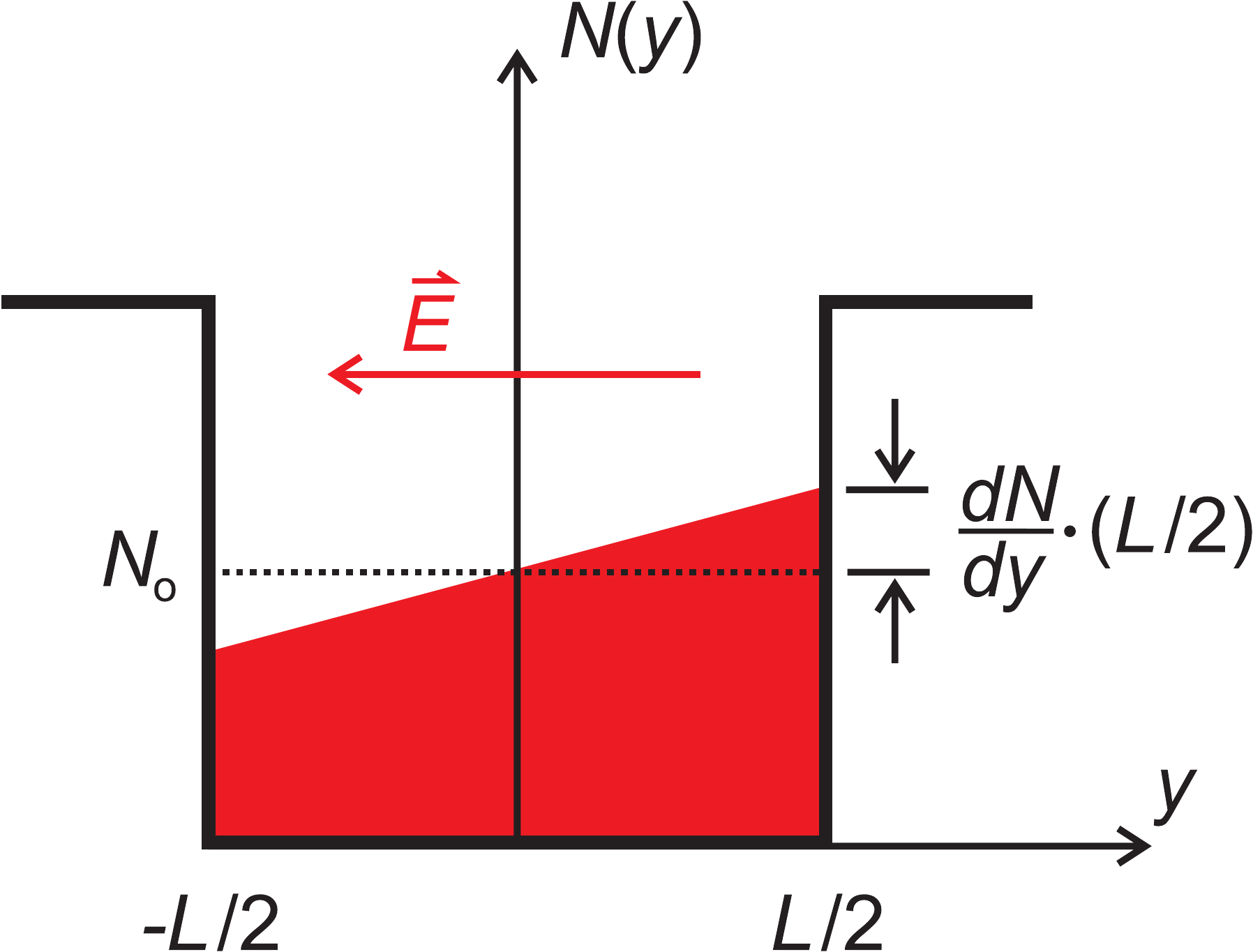}  
\vspace{0.5cm}
\caption{\small \sl Density gradient of non-interacting, classical electrons induced in a weakly confining nanostructure by a weak electric field, $\vec{E}$. $N$ is the carrier density, $N_{\circ}$ is the average carrier density, $y$ is the distance along the direction of the field, and $L$ is the box width.  \label{fig:Figure 4}}  
\end{center}  
\end{figure}  

The diffusion current $j_{\text{diff}}(t)$ responds to the instantaneous carrier density gradient created by the impulse and therefore will act to restore a zero gradient but go no further.  If the diffusion current rate is defined as 
\begin{equation}
a \equiv \frac{12 D}{L^{2}} \; \; ,
\end{equation}
then the total impulse response of the system for $R=1$ is
\begin{equation}
j(t) = j_{\circ}(t) - a \int_{0}^{t} j(t') dt' \; \; .
\end{equation}
The rationale for labelling the impulse response in the previous section $j_{\circ}(t)$ is now clear, as it does not include diffusion and is consequently an incomplete impulse response for our model system. Equation (41) is valid for any position $y$; by taking the derivative of both sides with respect to time, it can be rewritten as
\begin{equation}
\frac{dj(t)}{dt} = \frac{dj_{\circ}(t)}{dt} -aj(t) \; \; .
\end{equation}
Furthermore, since the conductivity is obtained via the Fourier transform of the impulse response, the differential equation (Eq. (42)) need not be solved.  Instead, taking the Fourier transform directly reveals
\begin{equation}
\tilde{\sigma} (\omega) = \tilde{\sigma}_{\circ} (\omega) \left( 1- \frac{1}{1-i \omega /a} \right) \; \; .
\end{equation}
Therefore, the total conductivity of our model system for $R=1$ is
\begin{equation}
\tilde{\sigma} ( \omega) = \frac{Ne^2 \tau ' /m^{*}}{1-i\omega \tau '} \left(1- \frac{1}{1-i \omega / a} \right) \; \; .
\end{equation}
\\ \indent In Eq. (44), we have made use of the equivalence between the collapsed Drude-Smith formula (Eq. (20) combined with Eq. (24)) and the results of our derivation from the previous section (Eq. (34)).  For $R=1$, $\tau ' = \left( \frac{1}{\tau} + \frac{2\text{v}_{\text{th}}}{L} \right) ^{-1}$, so for 100\% reflecting barriers, diffusion produces a conductivity that is almost identical to the truncated Drude-Smith formula (Eqs. (1) and (18)). Equation (44) is the modified Drude-Smith model referred to in the Abstract that describes, without free parameters, the conductivity of a weakly confined Drude gas of electrons inside a box with 100\% reflecting boundaries.
\\ \indent Recall the estimate for the length scale probed by a THz pulse given in the Introduction, $L_{\omega} \approx \sqrt{D/\omega}$.  For comparison, the box width at which $\omega/a=1$ is $L=\sqrt{12D/\omega}$.  In fact, the two estimates are based on the same physical principle, namely diffusion.  If a carrier can diffuse to a boundary during one period of the probing electric field then its transport will be confined over that time scale.  Similarly, if an electric field is applied in one direction over a sufficient time for carriers to reach the boundary and be confined, then a carrier density gradient will be established, resulting in a diffusion restoring current.
\\ \indent To compare our completed theoretical conductivity formula (Eq. (44)) with the results of the Monte Carlo simulation using no free parameters, we need to identify $a$.  The value of the diffusion coefficient depends on dimensionality if thermal motion is defined in multiple dimensions.  Our thermal velocity is one-dimensional, though, so we can use the one-dimensional diffusion coefficient to model the Monte Carlo simulation,
\begin{equation}
D=\frac{\tau' k_{B} T}{m^{*}} = \tau' \text{v}_{\text{th}}^{2} \; \; .
\end{equation}
We note that we replace $\tau$ with $\tau '$ (see Eq. (24)) in the diffusion coefficient of the Drude gas when including weak confinement, since the boundaries influence the rate of diffusion that would be present in an unconfined system. It was demonstrated in the previous subsection that to a good approximation, backscattering off the boundaries results in the Drude model with a modified scattering time for $L>\text{v}_{\text{th}}\tau$.  Since the conductivity of the system in the absence of diffusion is just that of an unconfined Drude gas with a slightly different character, we treat the diffusion coefficient in the same way.  Finally, for 100\% reflecting barriers,
\begin{equation}
a = \frac{12}{t_{\circ}} \left( \frac{\tau}{t_{\circ} + 2\tau} \right)
\end{equation}
where, as before, $t_{\circ}=L/\text{v}_{\text{th}}$, and we have once more made use of Eq. (24).
\\ \indent In combination with Eqs. (24) and (46), Eq. (44) can be determined for all sets of Monte Carlo parameters with $R=1$.  In the following section, we show that very good agreement is obtained between our theoretical prediction and the Monte Carlo simulations.  The conductivity formula that we have derived based on the diffusion restoring current is also very similar to the truncated Drude-Smith conductivity formula.  The correspondence is such that we label Eq. (44) a modified Drude-Smith model.
\\ \indent The generalization of Eq. (44) to include $R \neq 1$ is difficult and has not been completely solved here.  At first the problem appears trivial.  A fraction of the impulse response (1-$R$) will pass through the barrier unobstructed.  The transmitted current will produce a shift in average particle position that does not result in a carrier density gradient.  Hence, diffusion will not counteract the shift and the system will have a finite DC conductivity.  Of course, the impulse response that does not pass through the barrier will establish a carrier density gradient that will be restored to zero by diffusion.  The complication is that for $R \neq 1$ the carriers can also diffuse forward through the partially-transmissive barrier.  Forward diffusion will act to restore the carrier density gradient to zero because there is a difference in carrier density of $L\frac{dN}{dy}$ across the barrier, but it will also produce an additional DC conductivity.  Therefore, the DC conductivity will be larger than $(1-R)Ne^{2}\tau'/m^{*}$ for $R < 1$.  It is not obvious to us how forward diffusion should be incorporated into Eq. (44). We therefore propose a phenomenological ansatz based on the truncated Drude-Smith model:
\begin{equation}
\tilde{\sigma} ( \omega) = \frac{Ne^2 \tau ' /m^{*}}{1-i\omega \tau '} \left(1- \frac{c(R)}{1-i \omega / a}  \right) \; \; .
\end{equation}
The parameter $c(R)$ is expected to be less dependent on the box size than $c$ in the Drude-Smith model, but its precise dependence on $R$ and $L$ is not explored here. Equation (47) is the extension of the modified Drude-Smith model (Eq. (44)) promised in the Abstract that includes a single fit parameter to describe weakly confined charge carriers in a nanoscale box with boundary reflection probability $R\leq$1. We explore its ability to describe the results of the Monte Carlo simulations in the following sections.

\section{Results and Discussion}

Here, we show the conductivities derived from the Monte Carlo simulation and the modified Drude-Smith model for the same input parameters. For $R=1$ the model is complete (Eqs. (24), (44), and (46)) and we can compare the modified Drude-Smith model to the Monte Carlo simulations using no free parameters.  The parameters used for the Monte Carlo simulations shown in Fig. 5 were v$_{\text{th}}=2 \times 10^{5}$\,m/s, $E_{\circ}=1$\,kV/cm, $\tau =30$\,fs, and $R=1$.  These parameters allow the modified Drude-Smith model given by Eq. (44) to be determined exactly.  The comparisons for box sizes ranging from $L=20\text{v}_{\text{th}}\tau$\,=\,120\,nm to $L=\text{v}_{\text{th}}\tau$\,=\,6\,nm are shown in Fig. 5 (a)-(f).

\begin{figure}  
\begin{center}  
\includegraphics [width=85mm]{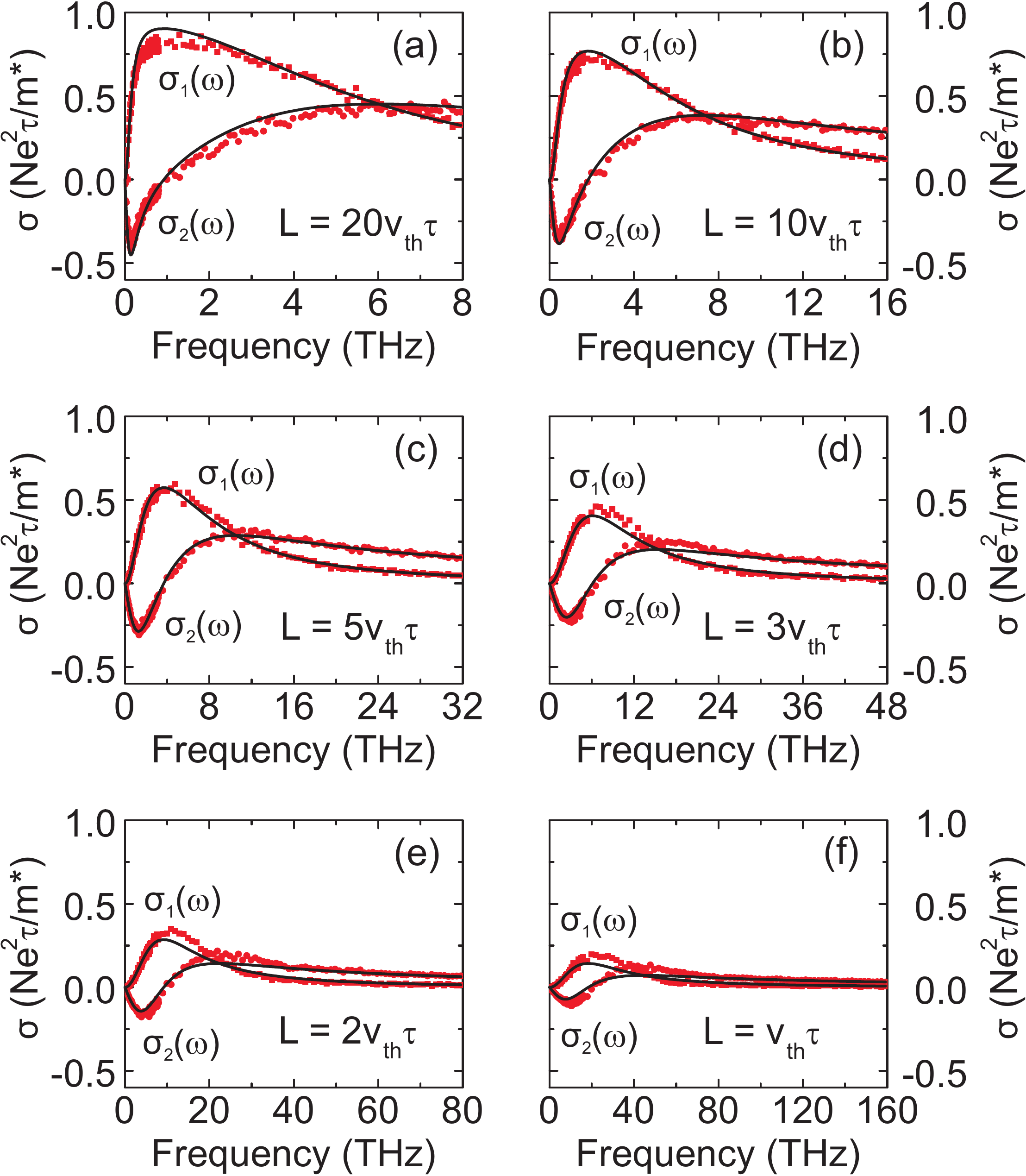}  
\vspace{0.5cm}
\caption{\small \sl Monte Carlo simulations of nanoparticles with 100\% reflecting barriers for a range of nanoparticle sizes.  The complex conductivities obtained from the Monte Carlo simulation for (a) $L=20\text{v}_{\text{th}}\tau$\,=\,120\,nm, (b) $L=10\text{v}_{\text{th}}\tau$\,=\,60\,nm, (c) $L=5\text{v}_{\text{th}}\tau$\,=\,30\,nm, (d) $L=3\text{v}_{\text{th}}\tau$\,=\,18\,nm, (e) $L=2\text{v}_{\text{th}}\tau$\,=\,12\,nm, and (f) $L=\text{v}_{\text{th}}\tau$\,=\,6\,nm are compared to the modified Drude-Smith model with $R$=1 (Eq. (44)).  Red squares denote the simulated $\sigma_{1}$, red circles denote the simulated $\sigma_{2}$ and black lines correspond to the modified Drude-Smith model.  No fit parameters are necessary for the modified Drude-Smith model because it is exactly determined by the input parameters of the Monte Carlo simulation.  For all simulations, $R=1$, $\text{v}_{\text{th}} = 2 \times 10^{5}$\,m/s, $\tau=30$\,fs, $m^{*}=0.26m_{e}$, and $E_{o}=1$\,kV/cm.\label{fig:Figure 5}}  
\end{center}  
\end{figure} 

\begin{figure}
\begin{center}  
\includegraphics [width=85mm]{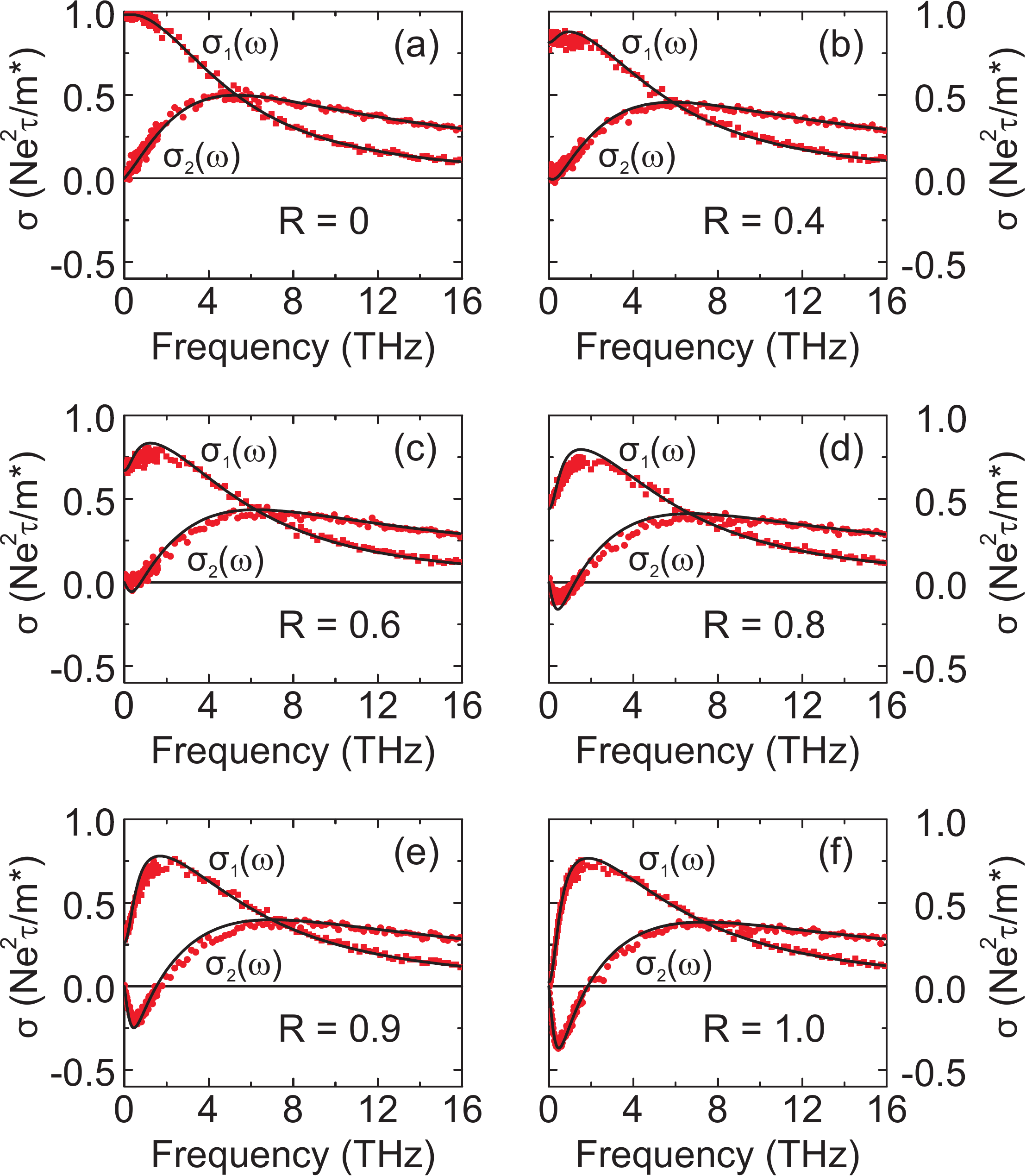}  
\vspace{0.5cm}
\caption{\small \sl Monte Carlo simulations of nanoparticles of width $L=10\text{v}_{\text{th}}\tau=60$\,nm for a range of barrier reflectivities.  Conductivities are shown for (a) $R=0$, (b) $R=0.4$, (c) $R=0.6$, (d) $R=0.8$, (e) $R=0.9$, and (f) $R=1.0$.  Red squares represent $\sigma_{1}$ and red circles represent $\sigma_{2}$.  Black lines are fits to the data by the modified Drude-Smith model for $R\leq$1 (Eq. (47)).  The lone fit parameter is $c(R)$, which is shown as a function of $R$ in Fig. 7.  The simulation parameters are $\text{v}_{\text{th}} = 2 \times 10^{5}$\,m/s, $\tau=30$\,fs, $m^{*}=0.26m_{e}$, and $E_{o}=1$\,kV/cm for all cases.\label{fig:Figure 6}}  
\end{center}  
\end{figure}  

The modified Drude-Smith model provides a remarkably good representation of the simulation data over the entire range of box sizes and over the full bandwidth of the simulations.  For small box sizes ($L=2\text{v}_{\text{th}}\tau$ in Fig.\,5(e), $L=\text{v}_{\text{th}}\tau$ in Fig. 5(f)) the modified Drude-Smith model deviates slightly from the simulation, possibly because the collapsed Drude-Smith model (Eqs. (20) and (24)) does not perfectly reproduce the conductivity described by Eq. (34) in this regime, as shown in Fig.\,3(e) and Fig.\,3(f). Scaling the $L=\text{v}_{\text{th}}\tau$ conductivity predicted by Eq. (44)  up by a factor of about 20\%, however, produces a good fit (not shown).  

\begin{figure}  
\begin{center}  
\includegraphics [width=60mm]{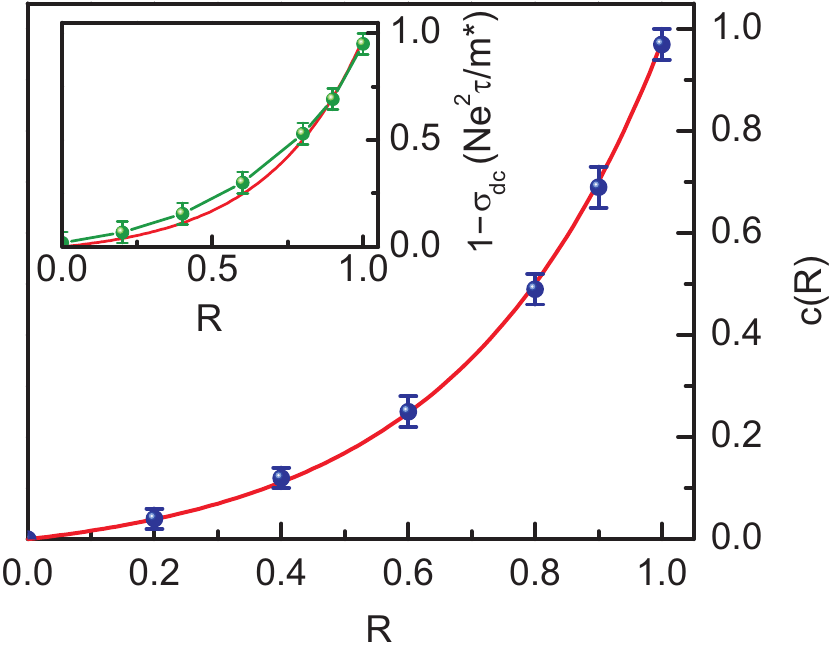}  
\vspace{0.5cm}
\caption{\small \sl Fit parameters $c(R)$ used to match the modified Drude-Smith model (Eq. (47)) to the Monte Carlo simulations in Fig. 6 ($L=10\text{v}_{\text{th}}\tau=60$\,nm, $R$ ranging from 0 to 1). Navy blue points show $c(R)$ values, while the red curve is an exponential fit to the points. Inset: Suppression of DC conductivities extracted from the Monte Carlo simulations, shown as green points with a connecting line (error bars are defined based on the scatter of points of the Monte Carlo conductivity). The exponential fit to the $c(R)$ data is shown as a red curve for comparison.\label{fig:Figure 7}}  
\end{center}  
\end{figure}  

On the other hand, the modified Drude-Smith model is not complete for $R \neq 1$, though we have proposed an ansatz, shown in Eq. (47). The functional form of the new parameter $c(R)$ is unknown, but since all the parameters of the Monte Carlo simulation are known, we can fit the $R$ dependence of the simulation using only $c(R)$.  We note that $c(R)$ is the only parameter that changes in the fitting, since the incorporation of $R$ into $a$ is easily accomplished via Eq. (24), Eq. (40), and Eq. (45).
\\ \indent Monte Carlo simulations were run with the parameters v$_{\text{th}}=2 \times 10^{5}$\,m/s, $E_{\text{o}}=1$\,kV/cm, $\tau =30$\,fs, $L=10 \text{v}_{\text{th}}\tau$, and $R$ ranging from 0 to 1.  The results are shown in Fig. 6(a)-(f).  Equation (47) provides good fits to the data for all simulations via the single fitting parameter $c(R)$.
\\ \indent The functional form of $c(R)$ was not derived in the previous section, so an estimate is made here based on the fitting results in Fig. 6. The $c(R)$ best-fit parameters used in Fig. 6(a)-(f) are shown in Fig. 7.  Within error, they follow the relation
\begin{equation}
c(R) = A \left( e^{BR} -1 \right)
\end{equation}
for $L=10\text{v}_{\text{th}}\tau$.  For the exponential fit in Fig. 7 (red curve), $A=0.045 \pm 0.002 $ and $B=3.12 \pm 0.05$.  The exponential dependence of $c(R)$ on $R$ suggests that its identity should be exactly solvable, but that step lies in the realm of future work. It is also possible that $c(R)$ has some dependence on $L$ (e.g. as shown in Ref. 76) but this requires further study. The inset of Fig. 7 shows the suppression of the Monte Carlo DC conductivities as a function of $R$. This tracks $c(R)$ and the exponential fit to $c(R)$ is shown for comparison. We note that the nonlinear progression of the DC conductivity with $R$ was anticipated based on forward diffusion through the barrier, since without forward diffusion the dependence would be close to linear.

\section{Modeling real systems}

Although the truncated Drude-Smith model has been widely applied to experimental data, interpreting the resulting fits has sometimes proven challenging. Even when it is treated as a purely phenomenological formula, problems can still arise. N\u{e}mec \textit{et al}. highlighted this issue by comparing the truncated Drude-Smith model to their Monte Carlo simulations over the frequency range of typical THz spectroscopy measurements.\cite{nemec2009} They demonstrated that the Drude-Smith fit parameters $\tau_{\text{\tiny{DS}}}$ and $c$ (see Eq. (1)) actually exhibit a complicated dependence on nanoparticle size, nanoparticle boundary reflectivity, and carrier mean free path. Moreover, they also found that the truncated Drude-Smith model was not in general capable of fitting their simulations over a broader spectral range.\cite{nemec2009} Conversely, Cooke \textit{et al}. showed that the truncated Drude-Smith model is capable of fitting the conductivities measured by ultrabroadband THz spectroscopy of silicon nanocrystal films.\cite{cooke2012a} In the following, we explain how the modified Drude-Smith model for $R$\,=\,1 (Eq. (44)) resolves this apparent discrepancy. We further provide recommendations for how and when to use the modified Drude-Smith model to describe experimental THz spectroscopy data.
\\ \indent We first consider the complex conductivity from our Monte Carlo simulations for $L=10\text{v}_{\text{th}}\tau$ and $R=$1, with $\tau$\,=\,30\,fs (Fig. 8(a), red squares and circles). The plot is divided into two frequency regimes: (i) 0\,-\,1.65\,THz, for comparison with conventional THz spectroscopy experiments, and (ii) 1.65\,-\,16\,THz. The modified Drude-Smith model (Eq. (44), solid black curves) accurately reproduces the simulated complex conductivity with no free parameters over the entire frequency range. In contrast, even with three free parameters ($\tau_{\text{\tiny{DS}}}$, $c$, and $N_{\text{\tiny{DS}}}$, which replaces $N$ in Eq. (1) and acts as a scaling factor) the truncated Drude-Smith model cannot fit the simulation results over the same bandwidth. The green dashed lines show a truncated Drude-Smith fit to the simulation data in frequency region (i). It not only fails to fit the simulation data in region (ii), but it also yields inaccurate fit parameters. Whereas the expected scattering time,
\begin{equation}
\tau '=\left( \frac{1}{\tau}+\frac{2}{t_{\circ}} \right)^{-1} \; \; ,
\end{equation}
is 25\,fs (for derivation see Eq. (24)), $\tau_{\text{\tiny{DS}}}$ for the green dashed curve is 120\,fs. This is between the scattering time of the simulation, $\tau$\,=\,30\,fs, and the average electron transit time across the box in the simulation, $t_{\circ}$\,=\,300\,fs, but close to neither. Additionally, $N_{\text{\tiny{DS}}}$\,=\,0.4$N$ for the green dashed curve rather than the correct value of $N_{\text{\tiny{DS}}}$\,=\,$N$, though $c$\,=\,-1 as expected. Fitting the Drude-Smith model instead to frequency range (ii) of the simulated conductivity gives the blue dotted curve in Fig.\,8(a), which clearly diverges from the simulated complex conductivity in region (i). In this case $\tau_{\text{\tiny{DS}}}$\,=\,48\,fs (i.e. it is closer to the real value but still incorrect), $c$\,= -1, and $N_{\text{\tiny{DS}}}$\,=\,$N$. Finally, if the Drude-Smith scattering rate is set to 25\,fs, the truncated Drude-Smith conductivity does not fit the simulated conductivity in either frequency range, as the peak of $\sigma_{1}$ blueshifts to ~6\,THz.
\\ \indent The reason that the truncated Drude-Smith model cannot fit the simulation data across the whole frequency range in Fig. 8(a) can be understood by comparing it to the modified Drude-Smith model. For $R$\,=\,1, the primary difference between the two formulas (Eq. (1) with $c$\,=\,-1 and Eq. (44), respectively) is that the modified Drude-Smith model contains two separate characteristic times. The first is $\tau'$, which is the inverse of the total scattering rate (see Eq. (49)). The second is the diffusion time, $t_{\text{diff}}$, where
\begin{equation}
t_{\text{diff}}=1/a=\frac{t_{\circ}}{12}\left( \frac{t_{\circ}+2\tau}{\tau} \right) \; \; ,
\end{equation}
so the modified Drude-Smith model for $R$=1 can be rewritten as
\begin{equation}
\tilde{\sigma} ( \omega) = \frac{Ne^2 \tau ' /m^{*}}{1-i\omega \tau '} \left(1- \frac{1}{1-i \omega t_{\text{diff}}} \right) \; \; .
\end{equation}
The truncated Drude-Smith model is effectively a special case of the modified Drude-Smith model in which $\tau_{\text{\tiny{DS}}}$\,=\,$\tau'$\,=\,$t_{\text{diff}}$. Hence, the failure of the Drude-Smith model to fit the Monte Carlo conductivity in Fig.\,8(a) occurs because this is not a good approximation for $L$\,=\,$10\text{v}_{\text{th}}\tau$. Indeed, calculating $t_{\text{diff}}$ and $\tau'$ for $L/(\text{v}_{\text{th}}\tau)$\,=\,$t_{\circ}/\tau$\,=\,10 reveals that $t_{\text{diff}}$\,=10$\tau$ and $\tau'$\,=\,$5\tau/6$. On the other hand, there is also a regime in which $\tau'$\,=\,$t_{\text{diff}}$ is a good approximation, and even a particular choice of $t_{\circ}/\tau$ for which it is exactly true. Setting Eq. (49) equal to Eq. (50), we find that this occurs for $t_{\circ}/\tau \approx$1.5.
\begin{figure}  
\begin{center}  
\includegraphics [width=60mm]{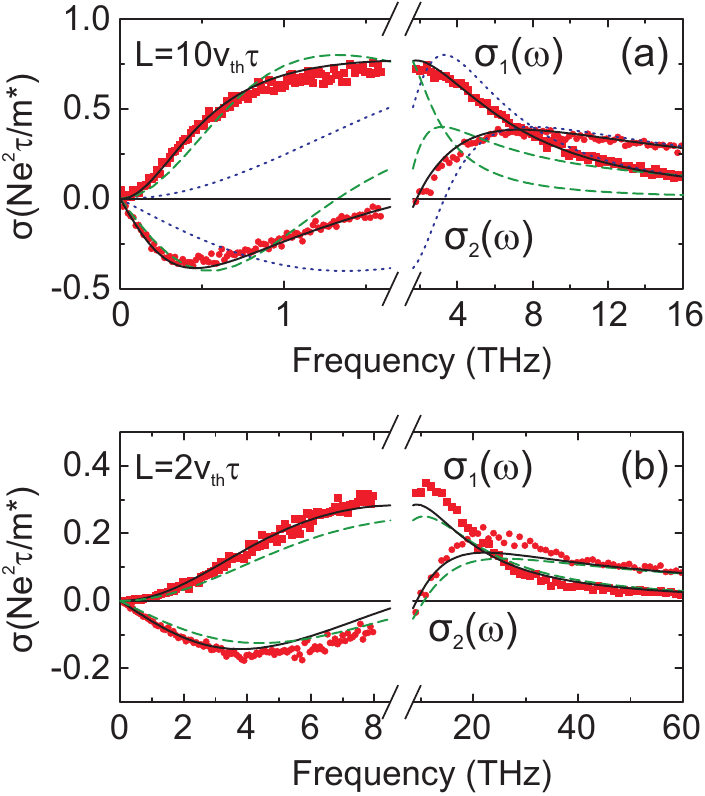}  
\vspace{0.5cm}
\caption{\small \sl Fitting conductivities with the modified Drude-Smith model vs. the truncated Drude-Smith model. (a) Monte Carlo simulation for a box with $L$\,=\,$10\text{v}_{\text{th}}\tau$\,=\,60\,nm and $R$\,=\,1 (red squares: $\sigma_{1}$; red circles: $\sigma_{2}$). The modified Drude-Smith model (solid black curves) is determined from the simulation parameters and has no free fit parameters. Two truncated Drude-Smith model fits to the simulated conductivity are shown: (i) one for low frequencies, below 1.65\,THz (green dashed curves), and (ii) one for high frequencies, above 1.65\,THz (blue dotted curves). The fit parameters for the green dashed curves are $c$\,=\,-1 and $\tau_{\text{\tiny{DS}}}$\,=\,120\,fs, and an additional scaling factor of 0.4 has been applied (i.e. $N_{\text{\tiny{DS}}}$\,=\,0.4$N$). The fit parameters for the blue dotted lines are $c$\,=\,-1 and $\tau_{\text{\tiny{DS}}}$\,=\,48\,fs, and no scaling factor has been applied. (b) Monte Carlo simulation for a box with $L$\,=\,$2\text{v}_{\text{th}}\tau$\,=\,12\,nm and $R$\,=\,1 (red squares: $\sigma_{1}$; red circles: $\sigma_{2}$). The modified Drude-Smith model (solid black curves) is again determined completely by the simulation parameters. Only one truncated Drude-Smith model curve is shown in this case (green dashed lines) because both the low and high frequency regions of the conductivity can be fit simultaneously, since $t_{\text{diff}}\approx\tau'$. The truncated Drude-Smith parameters are $c$\,=\,-1 and $\tau_{\text{\tiny{DS}}}$\,=\,$\tau'$\,=\,15\,fs, and no scaling factor is applied. For both simulations, $\tau$\,=\,30\,fs, $\text{v}_{\text{th}}$\,=\,2\,$\times$\,$10^{5}$\,m/s, $m^{*}$\,=\,$0.26m_{e}$, and $E_{\text{o}}$\,=\,1\,kV/cm. The scattering time $\tau$ in the axes refers to the intrinsic scattering time input into the simulations, i.e. $\tau$\,=\,30\,fs.\label{fig:Figure 8}}  
\end{center}  
\end{figure} 
\\ \indent Figure 8(b) shows the complex conductivity from a Monte Carlo simulation with $L$\,=\,$2\text{v}_{\text{th}}\tau$\,=\,12\,nm, $R$\,=\,1, and $\tau$\,=\,30\,fs (red squares and circles). Here, $t_{\circ}/\tau$\,=\,2, $t_{\text{diff}}$\,=\,2$\tau$/3, and $\tau'$\,=\,$\tau$/2, so the Drude-Smith approximation $t_{\text{diff}}\approx\tau'$ is more realistic. We have again split the dataset into two frequency regimes, but this time (i) corresponds to 0\,-\,8.5\,THz for comparison with the ultrabroadband spectroscopy of Cooke \textit{et al}.\cite{cooke2012a} and (ii) corresponds to 8.5\,-\,60\,THz. For this dataset, the modified Drude-Smith model (solid black curves) again provides a good representation of the complex conductivity over the entire frequency window with no free parameters, although its magnitude is $\sim$20\% too low, as was previously noted in Section IV. Additionally, in contrast to the previous dataset, the truncated Drude-Smith model also matches the simulation data reasonably well without free parameters. The green dashed line in Fig.\,8(b) is the Drude-Smith conductivity for $N_{\text{\tiny{DS}}}$\,=\,$N$, $c$\,=\,-1, and $\tau_{\text{\tiny{DS}}}$\,=\,15\,fs. These parameters are obtained from the normalization of the simulation data, the reflectivity of the box in the simulation, and $\tau_{\text{\tiny{DS}}}$\,=\,$\tau'$ (with $\tau'$ calculated using Eq.\,(49)), respectively. Thus, the truncated Drude-Smith model agrees with the Monte Carlo simulation for $L$\,=\,$2\text{v}_{\text{th}}\tau$, both in terms of the conductivity shape and parameter values. This explains the excellent agreement between the truncated Drude-Smith model and the experimental data over an ultrabroadband frequency range in Ref. 16, where the silicon nanocrystal sizes ranged from v$_{\text{th}}\tau$ to 3v$_{\text{th}}\tau$.
\\ \indent Since the modified Drude-Smith model does not suffer from the same limitations as the truncated Drude-Smith model (as evidenced by its good agreement with our Monte Carlo simulations), we expect it to fit all appropriate microscopic experimental conductivities over an ultrabroadband frequency range, not just those that satisfy $t_{\text{diff}}\approx\tau'$. The modified Drude-Smith model contains one additional parameter for fitting experimental data compared to the truncated Drude-Smith model, but as we've shown above this is necessary to accurately describe the two characteristic time scales of a weakly confined Drude gas of electrons. These two time scales, represented by $t_{\text{diff}}$ and $\tau'$ in the modified Drude-Smith model, affect the shape of the complex conductivity in different ways. It should therefore be possible to distinguish their respective influences on the conductivities measured by ultrabroadband THz spectroscopy. However, conventional THz spectroscopy encompasses a much narrower frequency window than that shown by Cooke \textit{et al}.,\cite{cooke2012a} with a typical range of $\sim$0.4\,-\,2.5\,THz . We anticipate that fitting the modified Drude-Smith model to conductivity data in such a limited frequency window will lead to fit parameter ambiguities in some cases, especially those in which the peak of $\sigma_{1}$ is far above or below the measurement bandwidth. Consequently, it is imperative that complementary techniques be used to independently corroborate the modified Drude-Smith model fit parameters. Experimentally, this can be done by, for example, imaging the morphology of the nanostructured sample or recording its DC conductivity via four-point-probe measurements.
\\ \indent The context of the modified Drude-Smith model should also be understood before applying it to experimental THz conductivities. We reiterate that it is a purely classical formula that was derived under the assumption of non-interacting particles. It should not be used to describe nanosystems exhibiting quantum confinement, as it is only valid when charge carriers can be approximated as a Drude gas. This defines a lower bound for the size of suitable nanosystems--roughly the carrier mean free path, the boundary between the microscopic and mesoscopic scales. It is therefore unimportant that the model deviates from our Monte Carlo simulations for $L$=v$_{\text{th}}\tau$ (Fig. 5(f)), because although both the approximation that the boundary scattering rate is roughly given by Poisson statistics (Fig. 3) and the approximation that the diffusion coefficient is Drude-like (Eq. (45)) break down, a real system would be quantum mechanical for $L<$v$_{\text{th}}\tau$ anyway. Drude-Smith-like conductivity signatures may also arise in some cases due to quantum confinement, tunneling, or transport effects,\cite{shimakawa2012,shimakawa2016,pushkarev2017} but such systems require an independent theoretical treatment and are beyond the scope of this work.
\\ \indent A second aspect which has not been incorporated into our model so far is the Coulomb interaction. As we described in the Introduction, the displacement of charge carriers in a nanoparticle by an external field results in a local depolarization field that will act to restore the equilibrium charge distribution. For non-percolated systems, the local nanoparticle conductivity thus contains a localized surface plasmon resonance that depends on carrier density and nanoparticle geometry. In isolated nanoparticles this resonance can be described by a Lorentz oscillator model,\cite{nienhuys2005,strait2009,parkinson2012,boland2016} and in systems of non-percolated nanoparticles an appropriate EMT can in some cases connect the macroscopic conductivity of the system to the microscopic conductivity inside the nanoparticles.\cite{mrozek2012,nemec2013,kuzel2014,zajac2014,nemec2015}. However, since the modified Drude-Smith conductivity represents the microscopic response of an entire nanoparticle (or even a nanoparticle network) and not the intrinsic conductivity of the material inside it, it is unclear whether it can be combined with conventional EMTs to account for depolarization fields. A revised formalism may be required, such as the approach of N\u{e}mec \textit{et al}.\cite{nemec2013} or Di Sia and Dallacasa.\cite{disia2011} Notably, simulations in Ref. 71 revealed that depolarization fields in percolated nanoparticle networks are minimal. Local-field corrections to the modified Drude-Smith model are therefore relevant mainly for non-percolated systems.
\\ \indent Finally, it is important to point out that while the conductivity measured by conventional THz spectroscopy can be affected by depolarization fields, it is actually the bare microscopic conductivity that is most relevant for next generation techniques like THz scanning tunneling microscopy\cite{cocker2013,cocker2016,yoshioka2016,jelic2017} and THz time-domain nanoscopy. \cite{chen2003,zhan2007,blanchard2014,eisele2014,mitrofanov2014,moon2015} These approaches can access the local conductivity inside single nanoparticles\cite{cocker2013,eisele2014} and operate beyond the restrictions of EMTs. As such, they may provide a fundamental test of the modified Drude-Smith model for weakly confined systems. 

\section{Conclusions}

In conclusion, we have derived an expression for the conductivity of a weakly confined Drude gas of classical, non-interacting electrons. Its predictions agree remarkably well with Monte Carlo simulations and require no free parameters for confining structures with 100\% reflecting barriers. A generalized formula is proposed for the case of partially transmissive barriers that also provides good agreement with simulations, but which requires one fitting parameter. The functional form of our model is very similar to the Drude-Smith formula that has been used to fit the THz conductivities of a wide variety of nanomaterials. However, we find that for our model system the characteristic conductivity shape is a consequence of diffusion and not carrier backscattering. For low frequencies a carrier density gradient is established that results in a diffusion current in the opposite direction of the drift current, reducing the net conductivity. This effect is intrinsically related to the probing length scale of a THz pulse--the distance a carrier diffuses in one period of the probing frequency. Our results do not negate the conventional interpretation of the Drude-Smith model in general, only for our model system, where classical particles are structurally confined. It is still possible that carrier backscattering is a valid explanation for the suppression of low-frequency conductivities in other types of systems. Nevertheless, for weakly confined charge carriers our modified Drude-Smith model provides a direct connection between THz conductivity and microscopic particle motion that will lead to new insight in future THz spectroscopy studies.

\begin{acknowledgments}
This work was supported by funding from the Natural Sciences and Engineering Research Council of Canada (NSERC), Canada Foundation for Innovation (CFI), Alberta Science and Research Investment Program (ASRIP), Alberta Innovates Technology Futures (AITF), and the iCORE Centre for Interdisciplinary Nanoscience (iCiNano). The authors thank D. G. Cooke for valuable discussions.
\end{acknowledgments}

\end{document}